\documentclass[%
 preprint,
superscriptaddress,
 amsmath,amssymb,
aps,
]{revtex4-2}

\usepackage{lmodern}
\usepackage{amsmath,amssymb,amsfonts}
\usepackage{graphicx}
\usepackage{booktabs}
\usepackage{bbm}
\usepackage{algorithm}
\usepackage{algpseudocode}
\usepackage{physics}
\usepackage{float}
\usepackage{placeins}
\usepackage{array}
\usepackage{wrapfig}

\usepackage{subcaption}   
\usepackage[utf8]{inputenc}
\usepackage[T1]{fontenc}
\begin{document}

\title{Optimization Strategies for Variational Quantum Algorithms in Noisy Landscapes}

\author{Vojtěch Novák}
\email{vojtech.novak.st1@vsb.cz}
\affiliation{
 Department of Computer Science, Faculty of Electrical Engineering and Computer science, VSB-Technical University of Ostrava, Ostrava, Czech Republic
}
\affiliation{
 IT4Innovations National Supercomputing Center, VSB - Technical University of Ostrava, 708 00 Ostrava, Czech Republic
}
\author{Ivan Zelinka}%
\affiliation{
 Dpt. of Informatics and Statistics, Marine Research Institute, Klaipeda University, Lithuania
}
\affiliation{
 IT4Innovations National Supercomputing Center, VSB - Technical University of Ostrava, 708 00 Ostrava, Czech Republic
}

\author{Václav Snášel}
\affiliation{
 Department of Computer Science, Faculty of Electrical Engineering and Computer science, VSB-Technical University of Ostrava, Ostrava, Czech Republic
}

\begin{abstract}
Variational Quantum Algorithms (VQAs) are a leading approach for near-term quantum computing but face major optimization challenges from noise, barren plateaus, and complex energy landscapes. We benchmarked more than fifty metaheuristic algorithms for the Variational Quantum Eigensolver (VQE) using a three-phase procedure: initial screening on the Ising model, scaling tests up to nine qubits, and convergence on a 192-parameter Hubbard model. Landscape visualizations revealed that smooth convex basins in noiseless settings become distorted and rugged under finite-shot sampling, explaining the failure of gradient-based local methods. Across models, CMA-ES and iL-SHADE consistently achieved the best performance, while Simulated Annealing (Cauchy), Harmony Search, and Symbiotic Organisms Search also showed robustness. In contrast, widely used optimizers such as PSO, GA, and standard DE variants degraded sharply with noise. These results identify a small set of resilient algorithms for noisy VQE and provide guidance for optimization strategies on near-term quantum devices.
\end{abstract}

\maketitle

\section{Introduction}

Quantum computing promises transformative advantages for problems intractable for classical computers, including simulating quantum systems, solving large-scale optimization, and advancing machine learning~\cite{cerezo2021variational, bharti2022noisy, endo2021hybrid, schuld2015introduction, biamonte2017quantum, cerezo2022challenges, di2023quantum}. Among the most promising near-term methods are Variational Quantum Algorithms (VQAs), which use quantum devices with classical optimization to tackle quantum chemistry, condensed matter, and combinatorial problems~\cite{cerezo2021variational, bharti2022noisy, endo2021hybrid}. VQAs are thus a leading paradigm for noisy intermediate-scale quantum (NISQ) devices.

The Variational Quantum Eigensolver (VQE), a cornerstone VQA, approximates ground-state energies of molecular systems beyond classical reach~\cite{peruzzo2014variational}. More broadly, VQAs apply to quantum phase transitions, machine learning with parameterized circuits, combinatorial optimization via the Quantum Approximate Optimization Algorithm (QAOA), and quantum metrology~\cite{cerezo2022challenges, farhi2014quantum, hadfield2019quantum}.

Despite this promise, VQAs face difficult optimization landscapes. The most severe issue is the \textit{barren plateau} phenomenon, where gradients vanish exponentially with qubit number, making optimization intractable~\cite{mcclean2018barren, larocca2024review, holmes2022barren, cerezo2021cost}. Local minima~\cite{bittel2021training, fontana2022nontrivial, anschuetz2022quantum, anschuetz2022critical, anschuetz2024unified} and inevitable \textit{sampling noise}~\cite{lavrijsen2020classical} further complicate training. Measurement noise scales as $1/\sqrt{N}$ with shot count and persists even with error correction, setting a fundamental precision limit. Small gradients in barren plateaus require exponentially many shots to overcome statistical fluctuations, making gradient-based optimization impractical.

This motivates \textit{metaheuristic algorithms}\textemdash evolutionary, swarm-based, or physics-inspired methods such as Differential Evolution (DE), Particle Swarm Optimization (PSO), and Covariance Matrix Adaptation Evolution Strategy (CMA-ES)\textemdash which rely less on local gradient estimates and may offer greater robustness to noise~\cite{faildo2023using, hansen2001cma, eberhart1995particle}. Beyond their empirical success in VQAs, the rationale is supported by landscape analysis in Sec.~\ref{sec:landscape}. Visualization shows that the 1D Ising model without external magnetic field and without noise yields a smooth, nearly convex basin, but once sampling noise is introduced spurious minima appear and gradients vanish. The Fermi-Hubbard model is even harsher: its rugged, multimodal, nonconvex surface with many local traps mirrors the challenges of strongly correlated systems. These cases illustrate that VQE landscapes under noise are often multimodal, distorted, and deceptive, undermining local descent and motivating global search strategies.

A useful parallel comes from the IEEE CEC competitions on evolutionary optimization \cite{GarcíaMartínez2017SinceCEC}, where benchmark suites include multimodal, rotated, shifted, and compositionally complex functions. Such functions closely resemble the stochastic, high-dimensional landscapes of VQAs, where sampling noise and measurement uncertainty create similarly treacherous terrains. Performance in CEC benchmarks has long been dominated by advanced Differential Evolution variants such as iL-SHADE, while CMA-ES has also achieved top results in several years. The same algorithms reappear as the most resilient methods for noisy VQE optimization, reinforcing their suitability for this domain.

Section~\ref{sec:vqa} reviews the structure and applications of variational quantum algorithms. Section~\ref{sec:barren} introduces the barren plateau phenomenon and related trainability challenges. Section~\ref{sec:models} presents the benchmark models: the Ising chain and the Fermi-Hubbard model. Section~\ref{sec:methodology} details cost evaluation, ansatz design, and noise modeling. Section~\ref{sec:experiment_design} describes our three-phase evaluation of over fifty metaheuristic algorithms. Section~\ref{sec:landscape} visualizes the optimization landscapes to interpret optimizer behavior. Section~\ref{sec:results} reports comparative performance across models and noise levels. We conclude with implications for VQE optimization strategies on near-term devices.

\section{Variational Quantum Algorithms}
\label{sec:vqa}
Variational quantum algorithms combine quantum hardware with classical optimization~\cite{cerezo2021variational}, designed for NISQ devices with limited qubits and depth~\cite{preskill2018quantum}. They are typically divided into problem-driven VQAs (e.g., VQE, QAOA) and data-driven quantum machine learning (QML) models~\cite{cerezo2021variational, schuld2015introduction, biamonte2017quantum}. We focus on VQAs, where parameterized quantum circuits (PQCs) generate trial states, expectation values are measured, and classical optimizers adjust parameters.

Representative applications include VQE for chemistry~\cite{peruzzo2014variational}, QAOA for optimization~\cite{farhi2014quantum}, and other simulation protocols. These have shown proof-of-principle success but remain limited by trainability.

\subsection{The Barren Plateau Phenomenon}
\label{sec:barren}
We consider an $n$-qubit system initialized in a quantum state $\rho$ and evolved by a parametrized quantum circuit $U(\boldsymbol{\theta})$, giving $\rho(\boldsymbol{\theta}) = U(\boldsymbol{\theta}) \rho U^{\dagger}(\boldsymbol{\theta})$. 
Given a measurement operator $O$, the loss function is defined as the expectation value
\begin{equation}
\ell_{\boldsymbol{\theta}}(\rho,O) = \mathrm{Tr}[\rho(\boldsymbol{\theta}) O].
\end{equation}
In practice, this loss is estimated from a finite number $N$ of measurement shots, so the estimator $\hat{\ell}_{\boldsymbol{\theta}}(\rho,O)$ has a sampling variance of order $1/\sqrt{N}$. 
Optimization consists of finding parameters $\boldsymbol{\theta}$ that minimize $\ell_{\boldsymbol{\theta}}(\rho,O)$.

A \textit{barren plateau} arises when the loss or its gradients become exponentially concentrated around their mean as the system size grows. 
Formally, the variance of a gradient component decays exponentially,
\begin{equation}
\mathrm{Var}_{\boldsymbol{\theta}}\!\left[\nabla_{\theta_\mu} \ell_{\boldsymbol{\theta}}(\rho,O)\right] \in \mathcal{O}\!\left(\frac{1}{b^n}\right),
\end{equation}
with $b>1$. 
Because the gradient signal shrinks faster than the statistical noise of order $1/\sqrt{N}$, resolving descent directions requires exponentially many measurements, making training inefficient.

Two main forms of barren plateaus are distinguished: \textit{probabilistic concentration with narrow gorges} and \textit{deterministic concentration}.

\textit{Probabilistic concentration and narrow gorges.}
In this case, the variance of the loss or its partial derivatives decays exponentially:
\begin{equation}
\mathrm{Var}_{\boldsymbol{\theta}}[\ell_{\boldsymbol{\theta}}(\rho,O)] \in \mathcal{O}\!\left(\frac{1}{b^n}\right), \qquad b>1.
\end{equation}
This means that for most parameters, the loss concentrates around its mean and typical gradients vanish. 
Chebyshev's inequality implies that the probability of deviations larger than $\delta$ is exponentially suppressed:
\begin{equation}
\Pr\!\left(|\ell_{\boldsymbol{\theta}}(\rho,O)-\mathbb{E}_{\theta}[\ell_{\theta}(\rho,O)]| \geq \delta\right) \in \mathcal{O}\!\left(\frac{1}{b^n}\right).
\end{equation}
Thus the landscape is nearly flat on average, yet exponentially narrow regions with larger gradients can still exist. 
These steep but rare directions are referred to as \textit{narrow gorges}. 
They provide potential optimization paths, but the chance of locating and exploiting them decreases exponentially with $n$.

\textit{Deterministic concentration.}
A stronger form of barren plateau occurs when the entire loss landscape is uniformly concentrated around a constant value $\ell_0$:
\begin{equation}
|\ell_{\boldsymbol{\theta}}(\rho,O)-\ell_0| \in \mathcal{O}\!\left(\frac{1}{b^n}\right), \qquad \forall \boldsymbol{\theta}.
\end{equation}
Here no valleys or gorges remain: the loss is flat across all parameters and gradients vanish globally. 
This represents the most severe case, where optimization by gradient-based methods becomes effectively impossible.

Plateaus stem from the curse of dimensionality in Hilbert space~\cite{cerezo2023does}. Overly expressive circuits explore vast subspaces randomly~\cite{larocca2022diagnosing}, while depolarizing noise drives states to the maximally mixed state, creating deterministic plateaus~\cite{wang2021noise}. The choice of initial state and observable further influences concentration effects~\cite{cerezo2021cost}. Barren plateaus differ from local minima: they result from statistical concentration rather than true landscape structure. Both intrinsic (circuit-based) and extrinsic (noise-induced) mechanisms exist.

These effects make scaling VQAs challenging, as exponential shot costs quickly exhaust resources. Connections between barren plateaus and classical simulability~\cite{cerezo2023does} suggest avoiding them may limit quantum advantage. This motivates noise-robust, population-based metaheuristics, which explore globally without requiring accurate gradients. Their potential to bypass barren plateaus underpins our benchmarking of optimization strategies for VQAs.

\section{Benchmark Models}
\label{sec:models}
The effectiveness of metaheuristic algorithms in VQE depends critically on the underlying optimization landscape's complexity, characterized by multimodality, noise sensitivity, and the presence of barren plateaus~\cite{mcclean2018barren,cerezo2021cost}. We strategically selected two benchmark models that exhibit these challenging features while remaining computationally tractable for systematic algorithm comparison. VQAs face significant challenges including local minima, barren plateaus, and the stochastic nature of quantum measurements~\cite{bittel2021training,arrasmith2021effect}, conditions where population-based metaheuristics have demonstrated particular advantage over gradient-based methods~\cite{faildo2023using,batched2024evolutionary}.

\subsection{The 1D Ising Model}

The 1D transverse-field Ising model without external magnetic field serves as our primary benchmark because it presents a well-characterized multimodal landscape that challenges gradient-based methods while remaining analytically tractable. The model Hamiltonian is defined as:
\begin{equation}
H = -\sum_{i=1}^{n-1} \sigma_z^{(i)} \sigma_z^{(i+1)}
\end{equation}
where $\sigma_z^{(i)}$ represents the Pauli-Z operator on qubit $i$. This system exhibits two degenerate ground states (all spins aligned up or down), creating a symmetric double-well potential that tests optimizers ability to locate global minima.

Local minima inherently arise from minimizing complex functions, and these problems worsen with the number of qubits since the Hilbert space grows exponentially~\cite{kuo2022hybrid}. The Ising model's multimodal landscape with exponentially scaling parameter space directly challenges the global search capabilities that distinguish evolutionary and swarm-based algorithms from local methods. Restricting the parameter space decreases the potential of getting stuck on barren plateaus and local minima~\cite{kuo2022hybrid}, making this model an ideal controlled environment for comparative analysis.

We systematically vary the system size from 3 to 9 qubits (12 to 36 parameters) to evaluate how algorithm performance degrades with increasing dimensionality—a critical test for practical VQE applications where evolutionary machine learning practices have proven effective in increasing the training capabilities of variational quantum algorithms~\cite{batched2024evolutionary}.

\subsection{The Fermi-Hubbard Model}

The Fermi-Hubbard model represents one of the benchmarks for testing quantum computational methods for condensed matter~\cite{stanisic2022observing}. This model provides our most challenging optimization landscape, representing strongly correlated electron systems that are beyond classical computational reach~\cite{leblanc2015solutions}. The 6-site Hubbard Hamiltonian combines kinetic and interaction terms:
\begin{equation}
H = -t\sum_{\langle i,j\rangle,s} (c^\dagger_{i,s}c_{j,s} + c^\dagger_{j,s}c_{i,s}) + U\sum_i n_{i,\uparrow}n_{i,\downarrow}
\end{equation}
where $t = U = 1$, and the Jordan-Wigner transformation maps fermionic operators to a 12-qubit system (192 parameters after ansatz parameterization).

The Hubbard model encodes some of the key physics of strongly-correlated electrons in materials, given the natural mapping of spin-orbitals to qubits~\cite{bauer2020quantum}. Unlike the Ising model's symmetric landscape, the energy error obtained from optimizing quantum circuits plateaus for larger numbers of parameters~\cite{alvertis2025classical}, creating a highly frustrated optimization surface where local minima scattered over the optimization landscape cause early termination~\cite{batched2024evolutionary}. VQE has successfully computed ground state energies for lattice sizes up to 12 sites, demonstrating its utility for strongly correlated systems~\cite{stanisic2022observing}, but the 192-parameter space presents significant optimization challenges.

This scale specifically targets the regime where evolutionary algorithms that rely on population diversity may overcome local convergence issues~\cite{lavrijsen2020classical}, yet quantum fluctuations can bring the system out of shallow local minima~\cite{apolloni1989quantum}—precisely the conditions where population-based metaheuristics should demonstrate advantage. We utilize exact diagonalization results to assess absolute optimizer accuracy, enabling precise quantification of metaheuristic performance in reaching the true ground state energy.

These models form a complementary benchmark suite: the Ising model provides controlled assessment of fundamental multimodal navigation capabilities, while the Hubbard model tests performance under realistic strongly-correlated system complexity. This two-tier approach allows systematic evaluation of when and why evolutionary and swarm-based algorithms outperform traditional gradient methods in quantum optimization landscapes.

\section{Methodology}
\label{sec:methodology}
The VQE algorithm represents a hybrid quantum-classical optimization framework where quantum hardware evaluates cost functions while classical optimizers navigate the parameter landscape~\cite{cerezo2021variational}. This section establishes the mathematical foundations of our approach, focusing on the critical role of sampling noise in realistic optimization scenarios.

\subsection{Cost Function Evaluation via Quantum Expectation Values}

VQE seeks to minimize the energy expectation value of a parametrized quantum state $|\psi(\boldsymbol{\theta})\rangle$ with respect to a target Hamiltonian $\hat{H}$~\cite{peruzzo2014variational}:
\begin{equation}
E(\boldsymbol{\theta}) = \langle\psi(\boldsymbol{\theta})|\hat{H}|\psi(\boldsymbol{\theta})\rangle
\end{equation}
Following the variational principle, this provides an upper bound to the true ground state energy: $E(\boldsymbol{\theta}) \geq E_0$~\cite{mcweeny2012methods}. 

For quantum implementation, the Hamiltonian decomposes into a linear combination of measurable Pauli operators:
\begin{equation}
\hat{H} = \sum_{k=0}^{4^n-1} w_k \hat{P}_k, \quad \hat{P}_k = \bigotimes_{j=0}^{n-1} \sigma_{k_j}
\end{equation}
where $\sigma_{k_j} \in \{I, X, Y, Z\}$ and $w_k$ are real coefficients. The expectation value becomes:
\begin{equation}
\langle\hat{H}\rangle = \sum_{k} w_k \langle\hat{P}_k\rangle
\end{equation}

Computational efficiency requires the number of non-zero Pauli terms to scale at most polynomially with the number of qubits~\cite{wecker2015progress}. For both the Ising and Hubbard models studied here, this condition is satisfied due to their local interaction structure.

\subsection{Parametrized Quantum Circuits and Ansatz Design}

The classical variational approach employs linear combinations of basis functions, but quantum circuits provide a fundamentally different parametrization strategy. We replace traditional basis expansions with hardware-efficient parametrized quantum circuits (PQCs) that can be implemented on near-term quantum devices~\cite{kandala2017hardware}.

\begin{figure}[htpb]
    \centering
    \includegraphics[width=1\linewidth]{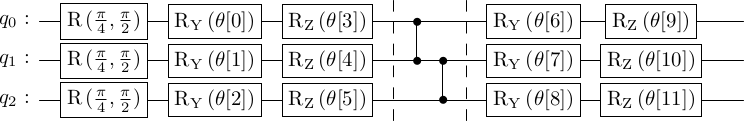}
    \caption{TwoLocal ansatz for three qubits with one repetition. The circuit begins with 
initialization $R\!\left(\tfrac{\pi}{4}, \tfrac{\pi}{2}\right)$ rotations on each qubit, 
followed by alternating single-qubit rotation layers ($R_Y$, $R_Z$) and entangling layers 
of controlled-$Z$ gates in a linear topology. Barriers separate the layers for clarity. 
This structure encodes the 1D Ising model without an external magnetic field and provides 
the parametrized state used within the VQE framework.}

    \label{fig:twolocal}
\end{figure}

Our primary ansatz follows the TwoLocal structure~\cite{sim2019expressibility}, alternating between single-qubit rotation layers and two-qubit entanglement layers:
\begin{equation}
|\psi(\boldsymbol{\theta})\rangle = \left[ \prod_{r=1}^{R} U_{\text{ent}} U_{\text{rot}}(\boldsymbol{\theta}^{(r)}) \right] U_{\text{rot}}(\boldsymbol{\theta}^{(0)}) |0\rangle^{\otimes n}
\end{equation}
where the final state $|\psi(\boldsymbol{\theta})\rangle$ is prepared by applying the circuit to the initial $n$-qubit state $|0\rangle^{\otimes n}$. The term $U_{\text{rot}}(\boldsymbol{\theta}^{(0)})$ is an initial layer of single-qubit rotations (e.g., RY gates). The main circuit consists of $R$ repetitions, where $R$ is the circuit depth. In each repetition, an entanglement layer $U_{\text{ent}}$ composed of two-qubit operations (controlled-Z gates) is applied, followed by a layer of parametrized single-qubit rotations $U_{\text{rot}}(\boldsymbol{\theta}^{(r)})$ (RY and RZ gates). The circuit is shown for 3 qubits in Fig. \ref{fig:twolocal}.

For both models, initialization begins with $R_Y(\pi/4)$ rotations on all qubits, creating an equal superposition state. For the 1D Ising model, this respects the system's $\mathbb{Z}_2$ symmetry, while for the Hubbard model it provides a balanced starting point for optimization. Linear entanglement connectivity is employed for both systems, with controlled-Z gates connecting adjacent qubits in the circuit layout.

The success of parametrized quantum circuits depends critically on expressibility (ability to span relevant Hilbert space regions) and trainability (optimization landscape properties)~\cite{sim2019expressibility,mcclean2018barren}. TwoLocal ansatz provide a favorable balance for NISQ applications, avoiding both underparameterization (insufficient expressibility) and overparameterization (trainability degradation due to barren plateaus).

Our conclusions may be ansatz-dependent. Beyond hardware-efficient TwoLocal circuits, problem-informed constructions such as the Hamiltonian Variational Ansatz (HVA/VHA) restrict the reachable state manifold by mirroring the target Hamiltonian’s locality and symmetries. This structure has been argued and demonstrated to improve trainability and to mitigate or avoid barren plateaus under appropriate parameterizations and initializations~\cite{PRXQuantum.1.020319,Park2024hamiltonian}. Recent VHA studies in noisy settings further show that sampling noise reshapes optimizer rankings: while gradient methods excel in noiseless regimes, population-based optimizers (e.g., CMA-ES) become more resilient under finite-shot noise, and physics-informed initializations reduce function-evaluation budgets~\cite{illesova2025numerical}.

Chemistry-inspired ansatz show similar sensitivities. For UCCSD, Trotterization and operator ordering choices can shift energies on chemically significant scales, implying that “Trotterized UCCSD’’ is not a single, well-defined object and may yield different outcomes across orderings and truncations~\cite{Grimsley2020TrotterizedUCCSD}. Generalized UCCSD (GUCCSD) extensions embedded in more advanced VQE schemes, such as state-averaged orbital-optimized VQE (SA-OO-VQE), have been proposed to simultaneously treat multiple near-degenerate states. These approaches demonstrate improved handling of excited states and quasi-diabatic representations, especially near conical intersections~\cite{Beseda_2024,Illesova2025TransformationFree}. Broad VQE surveys emphasize that expressibility, circuit depth, and optimizer behavior are tightly coupled to the chosen ansatz class, and that scalability is often limited by this interplay~\cite{Fedorov2022VQE}.

Our results with TwoLocal should be interpreted as a baseline. Stronger claims of generality require robustness checks across qualitatively different ansatz (e.g., HVA/VHA, UCCSD, and GUCCSD-based orbital-optimized variants) and noise models, ideally repeating the main experiments with (i) parameter-constrained HVA layers matched to model terms, and (ii) chemically motivated UCCSD/GUCCSD settings with controlled operator orderings. Where feasible, performance deltas relative to TwoLocal should be reported to isolate optimizer effects from ansatz-induced landscape changes.

\subsection{Sampling Noise and Statistical Uncertainty Analysis}

Unlike classical optimization where cost function evaluations are deterministic, quantum measurements introduce fundamental statistical uncertainty that significantly impacts optimizer performance~\cite{lavrijsen2020classical}. This sampling noise represents an unavoidable feature of quantum computation that we explicitly model and analyze.

Each Pauli operator measurement requires repeated quantum state preparation and readout. For $N$ measurement shots, the estimated expectation value follows:
\begin{equation}
\tilde{P}_k = \frac{1}{N} \sum_{i=1}^N m_i
\end{equation}
where $m_i \in \{-1, +1\}$ are individual measurement outcomes. Each outcome has variance
\begin{equation}
\mathrm{Var}[m_i] = 1 - \langle \hat{P}_k \rangle^2,
\end{equation}
so by the central limit theorem, for large $N$:
\begin{equation}
\tilde{P}_k \sim \mathcal{N}\!\left(\langle\hat{P}_k\rangle,\ \frac{1 - \langle \hat{P}_k \rangle^2}{N}\right), \qquad 
\mathrm{Var}[\tilde{P}_k] = \frac{1 - \langle \hat{P}_k \rangle^2}{N} \leq \frac{1}{N}.
\end{equation}

The total energy estimate variance depends on the shot allocation $\{N_k\}$ across terms:
\begin{equation}
\mathrm{Var}[\tilde{E}] = \sum_k w_k^2 \frac{1 - \langle \hat{P}_k \rangle^2}{N_k}.
\end{equation}
Assuming independent estimates with equal shot counts $N_k \equiv N$ and using $1-\langle \hat{P}_k\rangle^2 \leq 1$, this reduces to
\begin{equation}
\mathrm{Var}[\tilde{E}] \leq \frac{1}{N}\sum_k w_k^2.
\end{equation}

Applying Chebyshev's inequality provides probabilistic bounds on estimation accuracy~\cite{wainwright2019high}:
\begin{equation}
P(|\tilde{E} - E| \geq \epsilon) \leq \frac{\mathrm{Var}[\tilde{E}]}{\epsilon^2} \leq \frac{\sum_k w_k^2}{N\epsilon^2}.
\end{equation}
Tighter bounds can be obtained via Hoeffding or Bernstein inequalities, which exploit the $\pm 1$ boundedness of measurement outcomes:
\begin{equation}
P(|\tilde{E} - E| \geq \epsilon) \leq 2\exp\!\left(-\frac{2N\epsilon^2}{\sum_k w_k^2}\right).
\end{equation}

This establishes the fundamental trade-off between measurement precision and quantum resource consumption. To achieve energy estimation accuracy $\epsilon$ with confidence $(1-\delta)$, the required shot count scales as:
\begin{equation}
N \geq \frac{\sum_k w_k^2}{\delta \epsilon^2}.
\end{equation}
More generally, with a fixed shot budget $N_{\mathrm{tot}}$, the variance is minimized by allocating
\begin{equation}
N_k \propto |w_k|\sqrt{1 - \langle \hat{P}_k \rangle^2}, \qquad
\min \mathrm{Var}[\tilde E] = \frac{\left(\sum_k |w_k|\sqrt{1 - \langle \hat{P}_k \rangle^2}\right)^2}{N_{\mathrm{tot}}}.
\end{equation}

Sampling noise fundamentally alters the optimization problem structure. Instead of deterministic gradients, optimizers must navigate stochastic cost function landscapes where the noise level depends on both the circuit parameters and shot budget~\cite{wang2021noise}. This creates several challenges:

The effective noise floor is set by $\sigma_{\text{noise}} \approx \sqrt{\sum_k w_k^2}/\sqrt{N}$. Energy differences smaller than this noise level cannot be reliably distinguished, potentially masking optimization progress. For the systems studied, typical noise floors range from $10^{-2}$ (low shots) to $10^{-3}$ (high shots), requiring careful balance between measurement cost and optimization precision.

Gradient estimation becomes particularly challenging, as finite difference methods must overcome shot noise to detect parameter sensitivity. The parameter shift rule~\cite{mitarai2018quantum} provides unbiased gradient estimates but requires additional circuit evaluations, multiplying the effective shot cost.

We frame our study as an investigation of optimizer robustness to stochastic objective functions induced by finite-shot estimation. Shot noise is fundamental and persists as hardware improves, while modern error-suppression and mitigation can compress many coherent contributions toward effectively stochastic noise during expectation estimation~\cite{kim2023evidence,kim2023zne,cai2023qem,wallman2015rc,huggins2021efficient}. This choice isolates the optimizer--noise interface and aligns with recent demonstrations where scalable mitigation enables accurate expectation values at volumes beyond brute-force classical verification~\cite{kim2023evidence,kim2023zne}.

We explicitly acknowledge limits: coherent and correlated errors bias the cost landscape and can qualitatively change optimization. Theory and experiments show noise-induced barren plateaus and even noise-driven switches of the global minimizer, implying that algorithm rankings can differ under coherent or non-unital noise~\cite{wang2021nibp,cerezo2021cost,li2024noiseTransition}. Mitigation that tailors coherent noise to stochastic (e.g., randomized compiling), or combines with ZNE or data-driven regressors, can partially restore the stochastic regime but may change costs and sample budgets~\cite{wallman2015rc,kurita2022rczne,jiang2024gpr,cai2023qem}.

Results here should be read as lower bounds on robustness to statistical fluctuations. Rankings under coherent/systematic errors are hardware- and mitigation-dependent; future work can probe stability by repeating a subset of experiments with (i) randomized compiling on/off, and (ii) injected coherent over-rotations, reporting rank changes alongside shot budgets~\cite{wallman2015rc,kim2023zne}. Finally, measurement-cost reductions (grouping/low-rank factorization) can trade samples for circuit depth and interact with optimizer stochasticity~\cite{huggins2021efficient}.

\section{Experiments Design}
\label{sec:experiment_design}

In this study, we conducted numerical experiments to evaluate a diverse array of metaheuristic algorithms, categorized based on their inspiration or operational paradigm. The categories included bio-based, evolutionary-based, human-based, math-based, music-based, physics-based, swarm-based, and system-based algorithms. A detailed list of all optimizers is provided in Table~\ref{tab:optimizers_table}. For more details on each optimizer see Appendix \ref{sec:details} and settings used see Appendix \ref{appendix_params}.

Additionally, we performed initial experiments using gradient-based optimizers to assess their capability of reaching the global minimum. These experiments demonstrated that gradient-based optimizers often fail to converge to the global minimum, even with a relatively high tolerance of $10^{-1}$. The success rate of gradient optimizers was found to be low, which motivated the exploration of metaheuristic algorithms for more robust performance in VQE optimization tasks.

The experiments with meta-heuristic optimizers were divided into three phases due to the large number of optimizers under evaluation:
\begin{itemize}
    \item Phase 1: Initial selection
The first phase determined which optimizers were able to find the global minimum with a precision of $10^{-1}$. Each optimizer performed 5 independent runs on an Ising field model without an external magnetic field, using 5 qubits (corresponding to 20 parameters to optimize). If an optimizer reached the global minimum with the specified precision in at least one of the runs, it advanced to the second phase.
    \item Phase 2: Comparison of function evaluations
In the second phase, optimizers were compared based on their function evaluations (FEs). This was conducted on an Ising field model without an external magnetic field, with qubit counts ranging from 3 to 9. The function evaluations were calculated as the mean of 5 independent runs for each qubit configuration, up to the point where the optimizer reached the global minimum with a precision of $10^{-1}$. If an optimizer failed to reach the global minimum within the specified tolerance in any of the runs, its FEs were denoted as `\textemdash ' in the results table.
    \item Phase 3: Convergence on the Hubbard Model
In the third phase, convergence rates of the top-performing optimizers from the previous phases were analyzed on the Hubbard model with 192 parameters. Convergence curves were plotted using the mean function evaluations over 5 independent runs. Additionally, convergence was tested under two different shot settings for the estimator: a high-shot scenario with 5120 shots and a low-shot scenario with 64 shots.
\end{itemize}

\newpage

\begin{table}[htpb]
    \centering
    \caption{Table of used optimizers}
    \small
    \begin{tabular}{|p{2.5cm}|p{11cm}|}
        \hline
        \textbf{Category} & \textbf{Algorithms} \\
        \hline
        Bio-Based & BBOA \cite{bbobear2023}, SMA \cite{sma2020}, BBO \cite{bbo2008}, BMO \cite{bmo2018}, EOA \cite{eoa2015}, IWO \cite{iwo2006}, SBO \cite{sbo2017} \\
        & SOA \cite{soa2018}, SOS \cite{sos2014}, TPO \cite{tpo2018}, TSA \cite{tsa2020}, VCS \cite{vcs2015}, WHO \cite{who2019}, ABC \cite{ABC2009} \\
        \hline
        Evolutionary-Based & CMA-ES \cite{cmaes2001}, CRO \cite{cro2014}, EP \cite{yao1999evolutionary}, DE \cite{price2013differential}, GA \cite{mirjalili2019genetic}, FPA \cite{abdel2019flower}, MA \cite{neri2012memetic}, SHADE \cite{shade2013}, HyDE \cite{hyde2015} \\
        \hline
        Human-Based & BRO \cite{bro}, IBSO \cite{ibso}, CA \cite{ca}, CHIO \cite{chio}, FBIO \cite{fbio} \\
        & GSKA \cite{gska}, HBO \cite{hbo2020}, HCO \cite{hco2022}, ICA \cite{ica2007}, LCO \cite{lco2019}, QSA \cite{qsa} \\
        & SARO \cite{saro}, SPBO \cite{spbo2020}, SSDO \cite{ssdo2019}, TLO \cite{tlo2012}, TOA \cite{toa2021} \\
        \hline
        Math-Based & AOA \cite{aoa}, CEM \cite{cem2005}, CGO \cite{cgo2020}, CircleSA \cite{circlesa2022}, GBO \cite{gbo}, HC \cite{hc}, PSS \cite{pss}, RUN \cite{run}, SCA \cite{sca}, SHIO \cite{shio}, TS \cite{tabu} \\
        \hline
        Physics-Based & ASO \cite{aso2018}, ArchOA \cite{archoa2020}, CDO \cite{cdo2023}, EO \cite{eo2019}, EVO \cite{evo2022}, FLA \cite{fla}, HGSO \cite{hgso2019}, MVO \cite{mvo}, NRO \cite{nro2019}, RIME \cite{rime2023}, TWO \cite{two}, WDO \cite{wdo}, SA \cite{kirkpatrick1983optimization} \\
        \hline
        Swarm-Based & PSO \cite{eberhart1995particle}, iSOMA \cite{diep2020soma}, WOA \cite{WOA2016}, ALO \cite{ALO2015}, EHO \cite{EHO2015}, HHO \cite{HHO2019} \\
        \hline
    \end{tabular}
    \label{tab:optimizers_table}
\end{table}

All experiments were conducted on a machine with an Intel(R) Core(TM) i5-8400 CPU @ 2.80~GHz and 16~GB RAM. However, hardware specifications are not critical, as comparisons are made based on function evaluations rather than runtime. 

Unless explicitly stated otherwise, the estimator was set to use 5120 shots. The function evaluations were computed as the mean over five runs for each optimizer. 

The code for the experiments is publicly available as Jupyter notebooks at \url{https://github.com/VojtechNovak/VQA_metaheuristics} to ensure reproducibility of results and to allow further experimentation with optimizer hyperparameters. The implementation is based on stable Qiskit version~1 \cite{qiskit2024}. Most of the optimizers were obtained from the Python library \texttt{mealpy}, while CMA-ES was obtained from the \texttt{cma} module. Gradient-based optimizers were implemented using the Qiskit interface with \texttt{ scipy}, and the iL-SHADE algorithm was sourced from the \texttt{pyade} module.

\section{Optimization landscape visualization and analysis}
\label{sec:landscape}
To anticipate optimizer performance in VQE we first analyze the structure of the optimization landscapes. The methodology is straightforward: all variational parameters are fixed to the values obtained from a prior optimization, and only two selected parameters are varied across a two-dimensional grid. This two-dimensional slicing reveals local geometry around the optimum while remaining computationally simple. For the Ising model we study two parameter ranges, a global view with $\delta = 10$ that exposes the periodic structure and a local view with $\delta = 0.5$ that zooms into the basin of the optimum.

\begin{figure*}[t]
\centering
\begin{subfigure}[t]{0.45\textwidth}
\includegraphics[width=\linewidth]{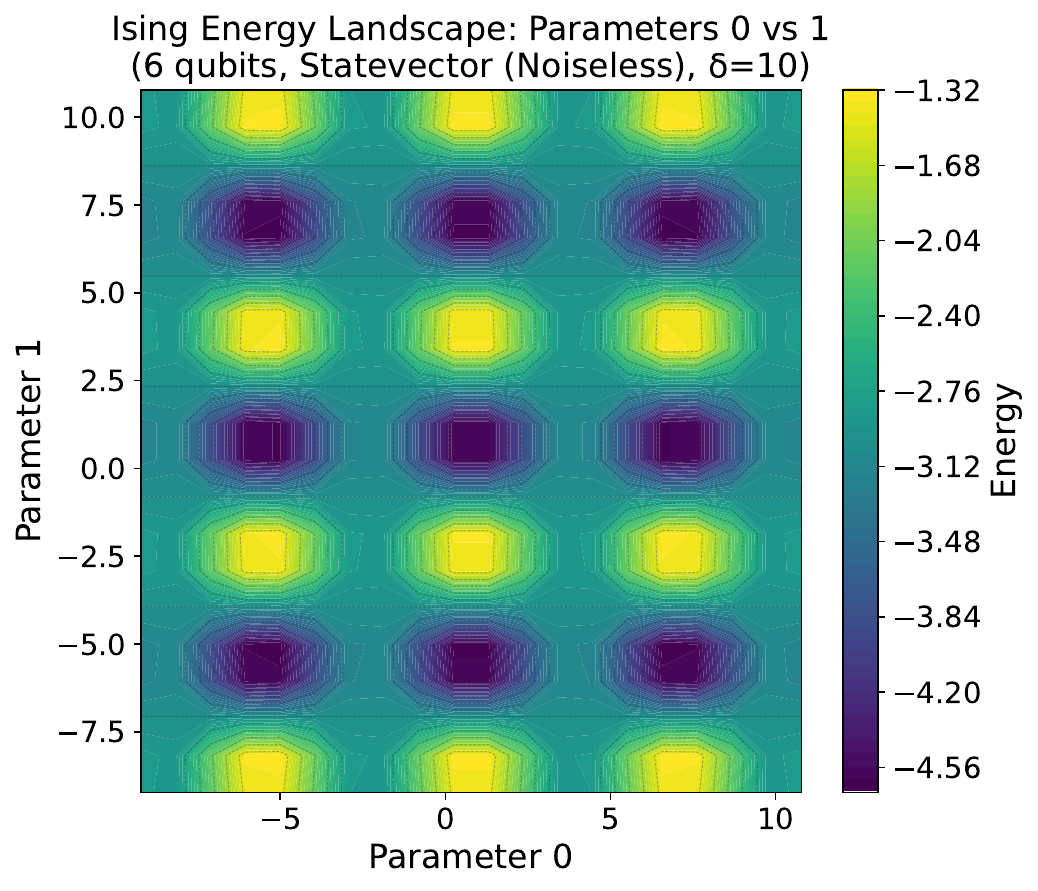}
\end{subfigure}
\begin{subfigure}[t]{0.45\textwidth}
\includegraphics[width=\linewidth]{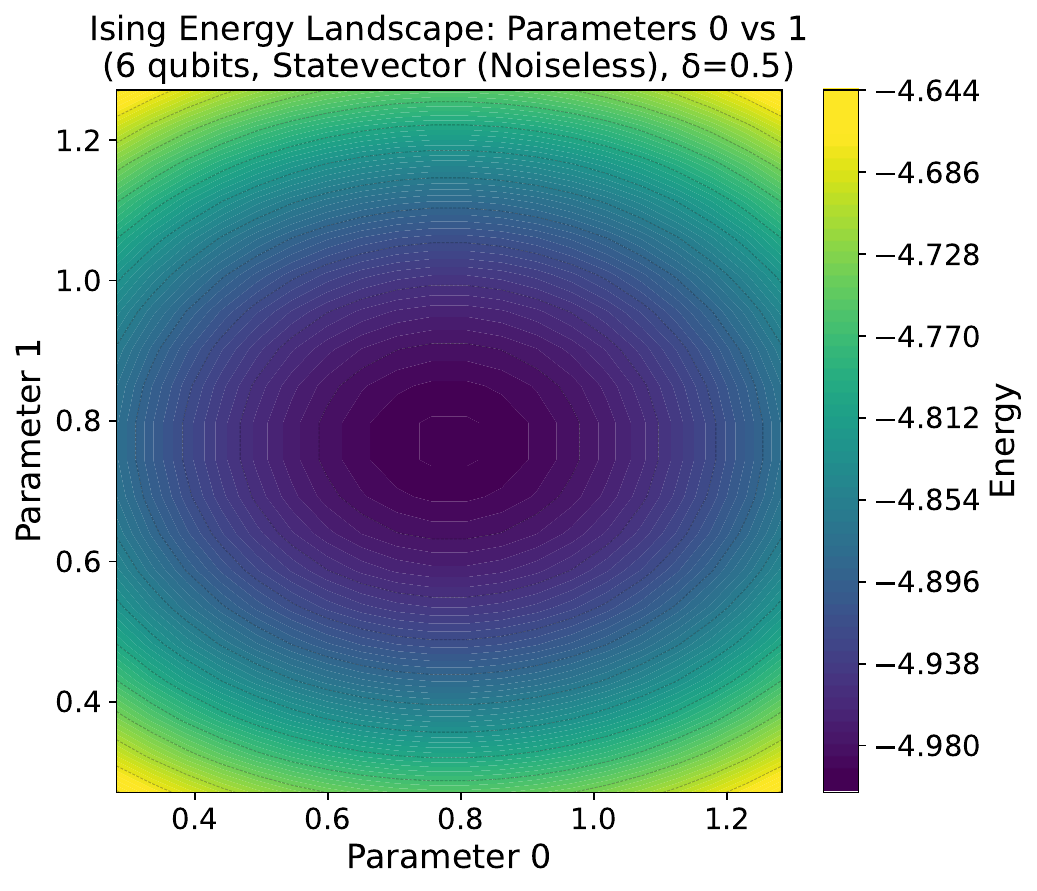}
\end{subfigure}\\[0.6em]

\begin{subfigure}[t]{0.45\textwidth}
\includegraphics[width=\linewidth]{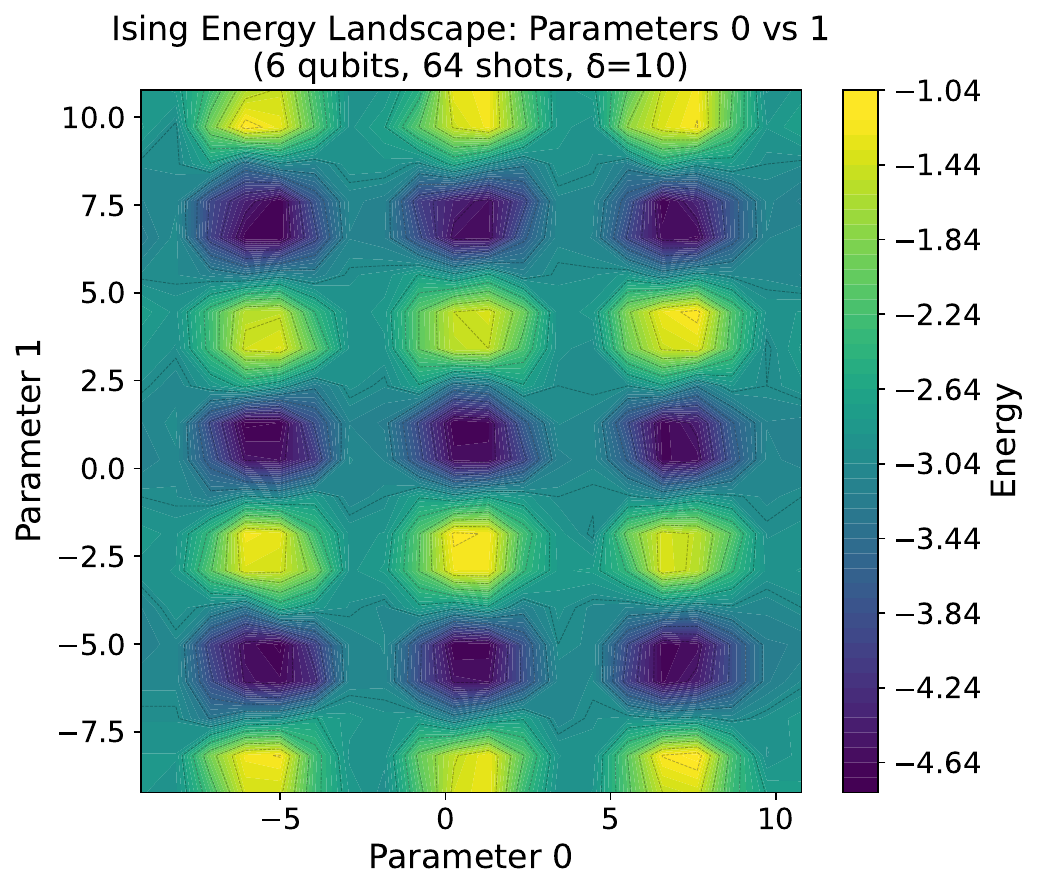}
\end{subfigure}
\begin{subfigure}[t]{0.45\textwidth}
\includegraphics[width=\linewidth]{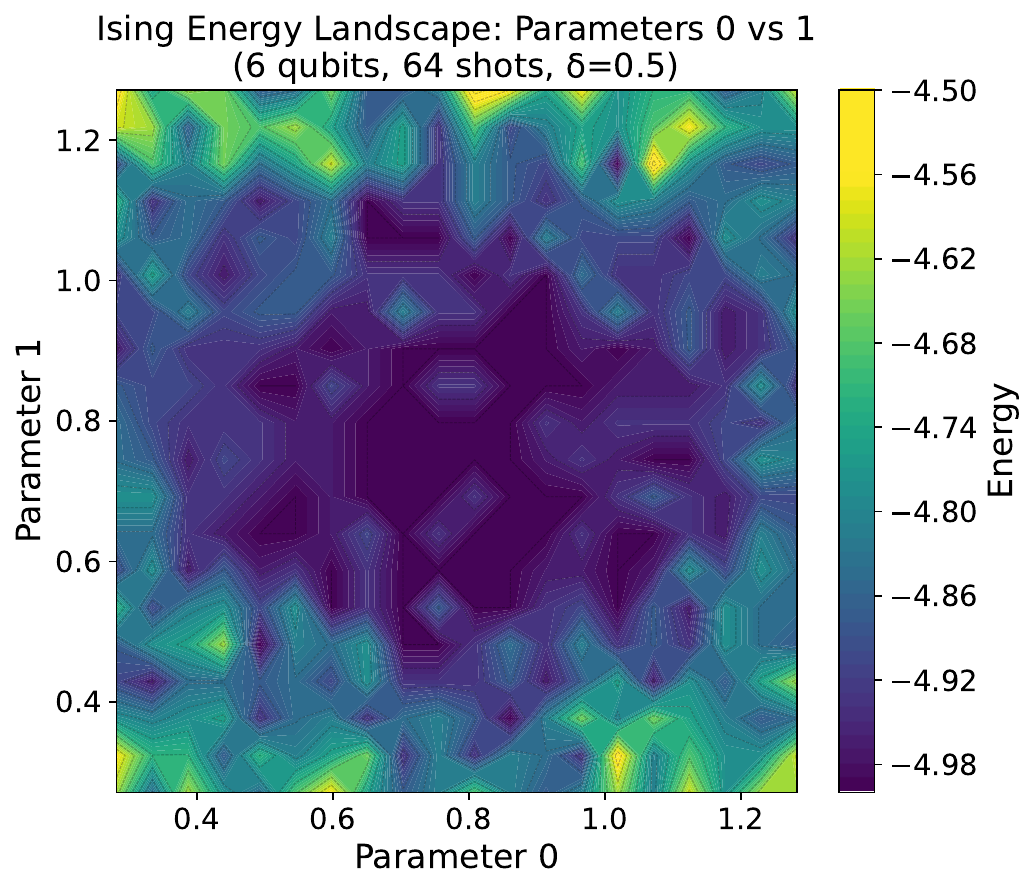}
\end{subfigure}\\[0.6em]

\begin{subfigure}[t]{0.45\textwidth}
\includegraphics[width=\linewidth]{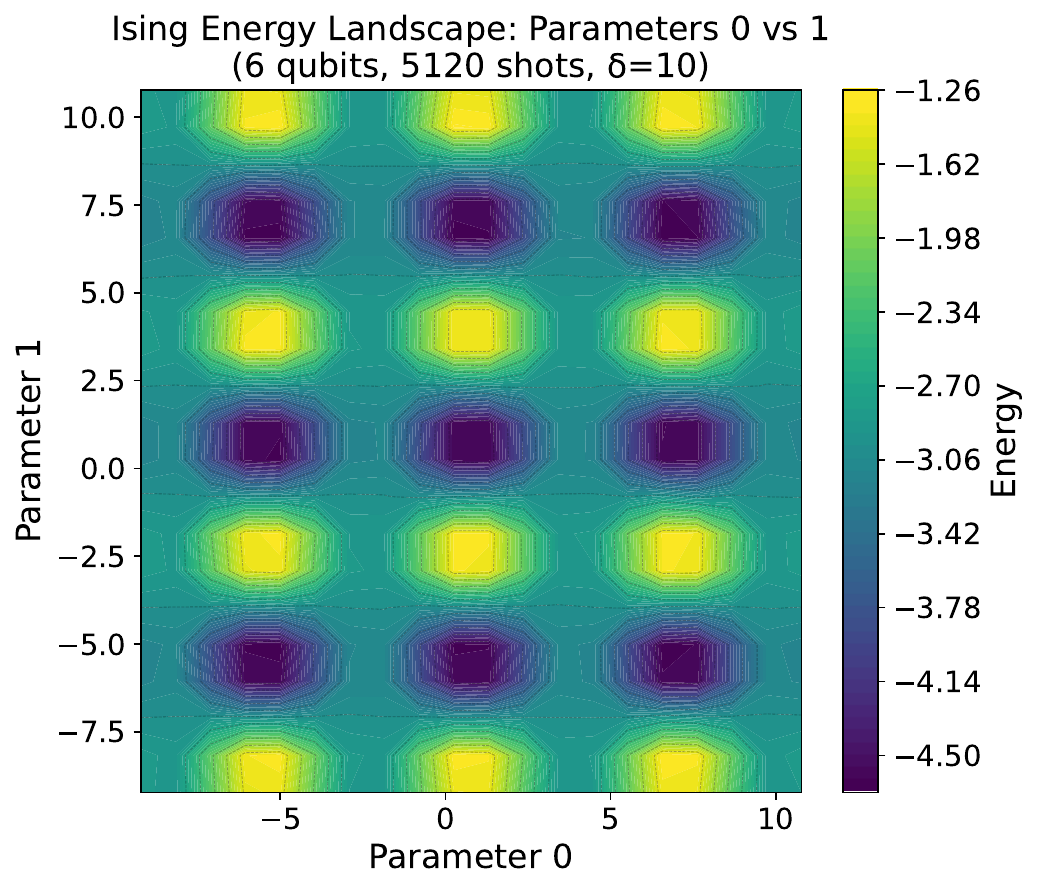}
\end{subfigure}
\begin{subfigure}[t]{0.45\textwidth}
\includegraphics[width=\linewidth]{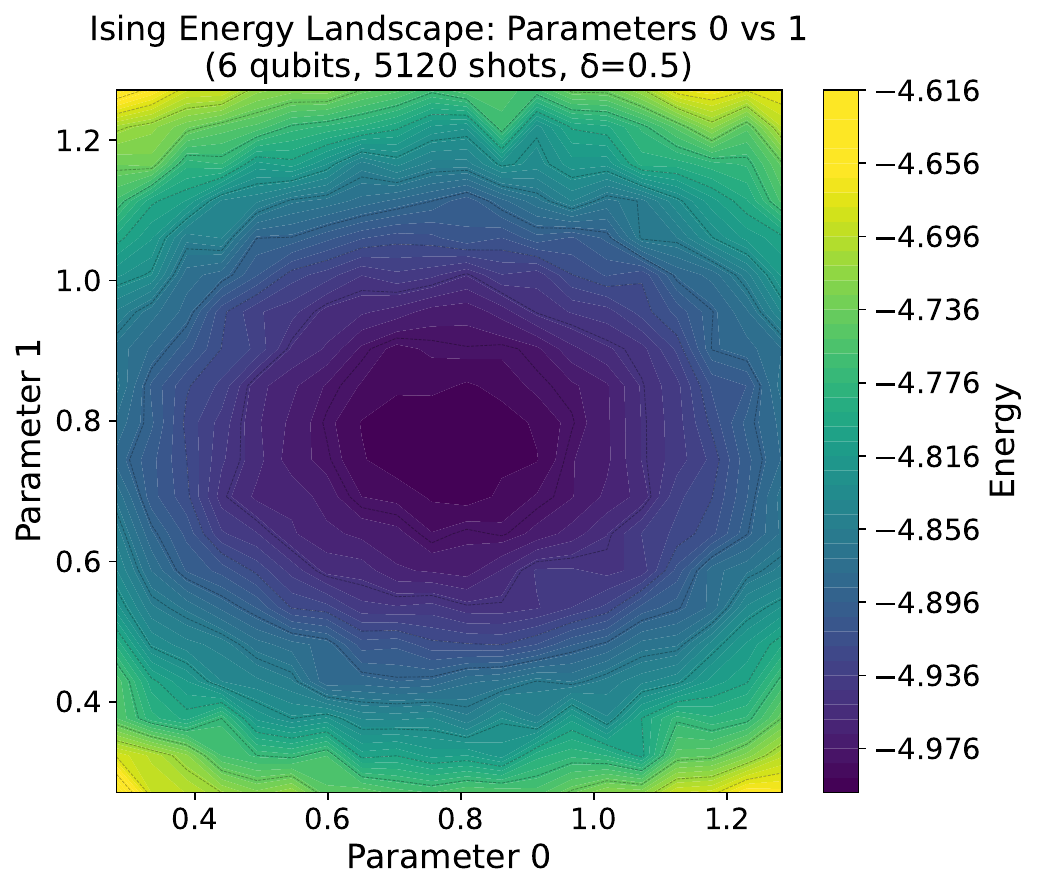}
\end{subfigure}
\caption{Energy landscapes, 6 qubits, showing landscape for parameters $\theta_0$ vs $\theta_1$.}
\label{fig:landscape-6q-p0vs1}
\end{figure*}

\begin{figure*}[t]
\centering
\begin{subfigure}[t]{0.45\textwidth}
\includegraphics[width=\linewidth]{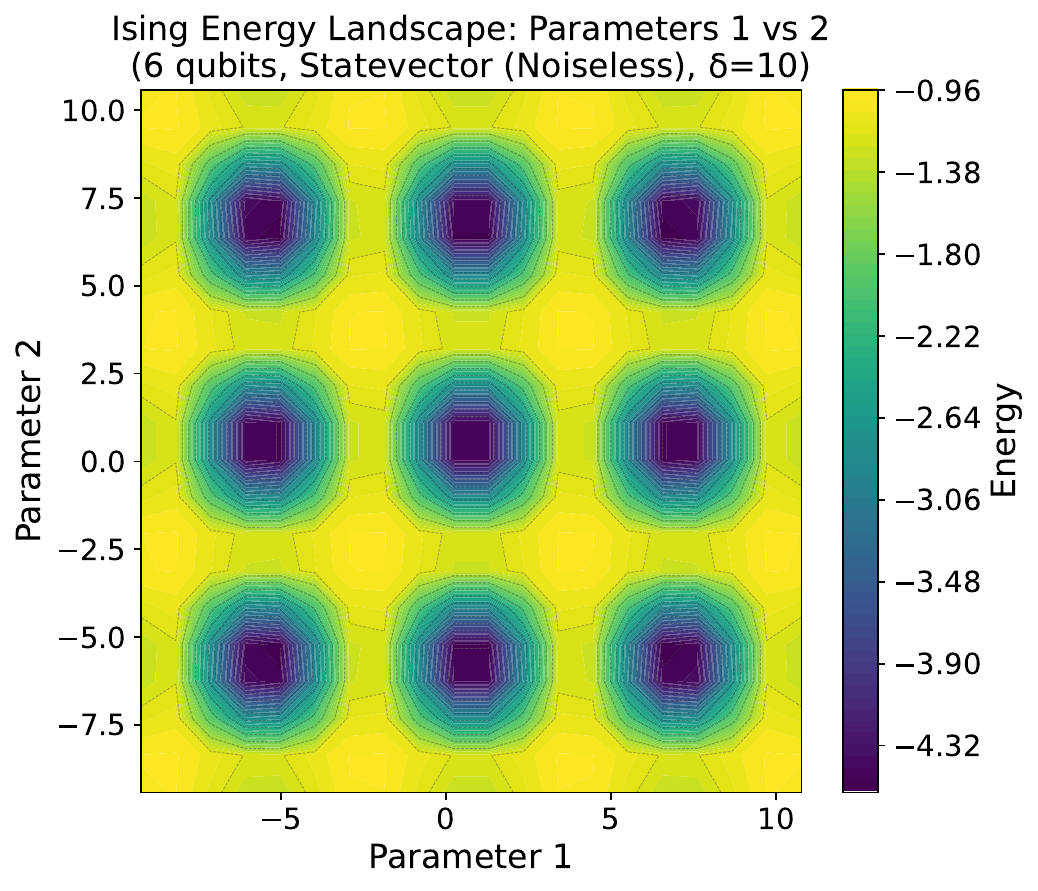}
\end{subfigure}
\begin{subfigure}[t]{0.45\textwidth}
\includegraphics[width=\linewidth]{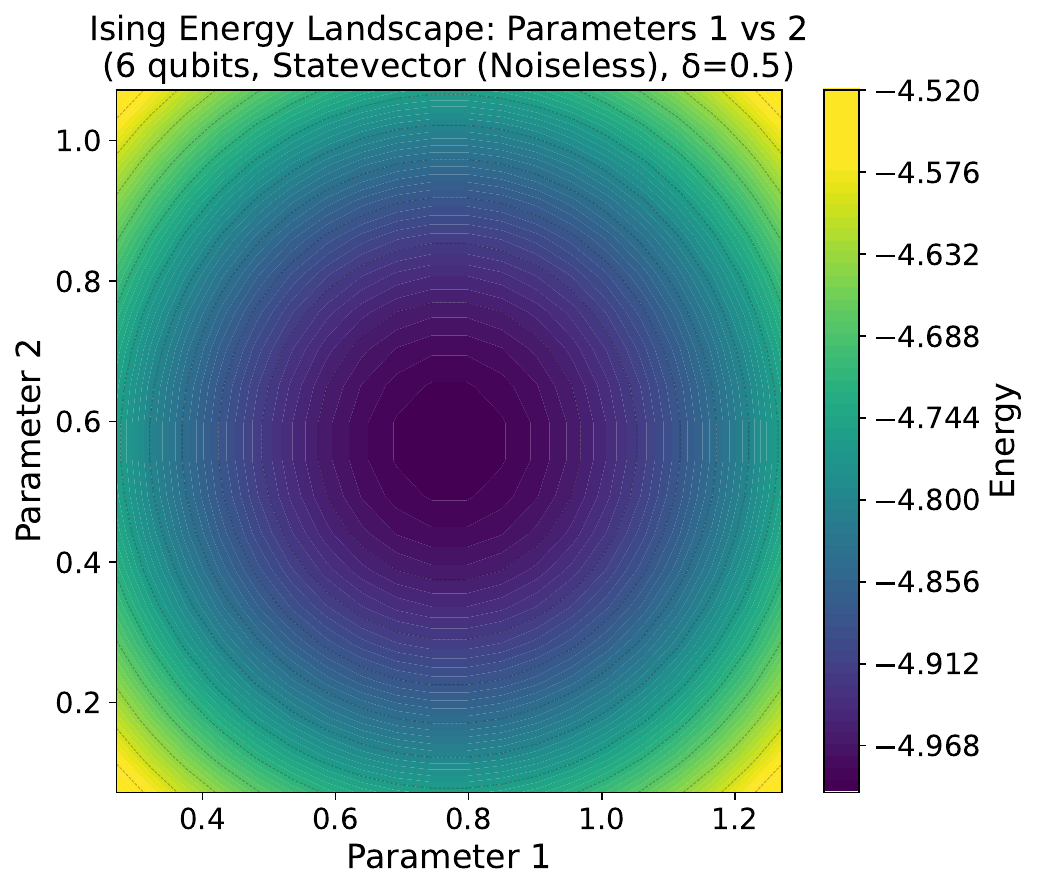}
\end{subfigure}\\[0.6em]

\begin{subfigure}[t]{0.45\textwidth}
\includegraphics[width=\linewidth]{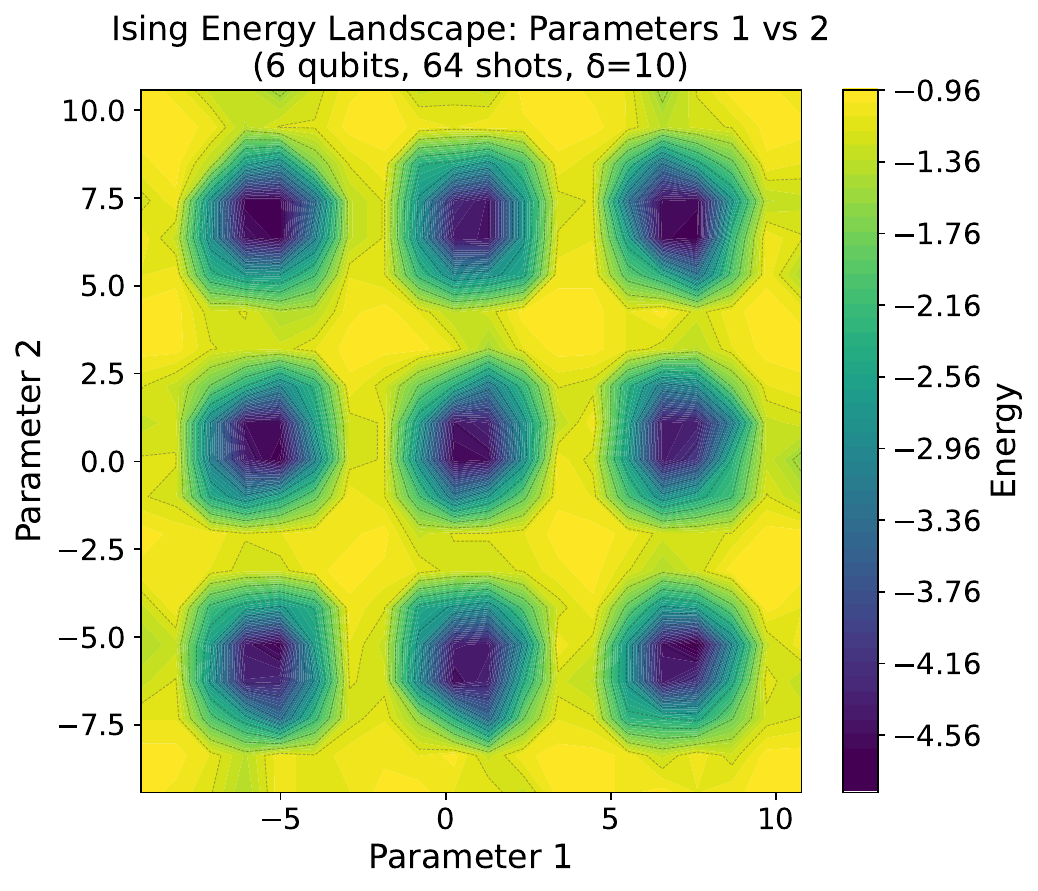}
\end{subfigure}
\begin{subfigure}[t]{0.45\textwidth}
\includegraphics[width=\linewidth]{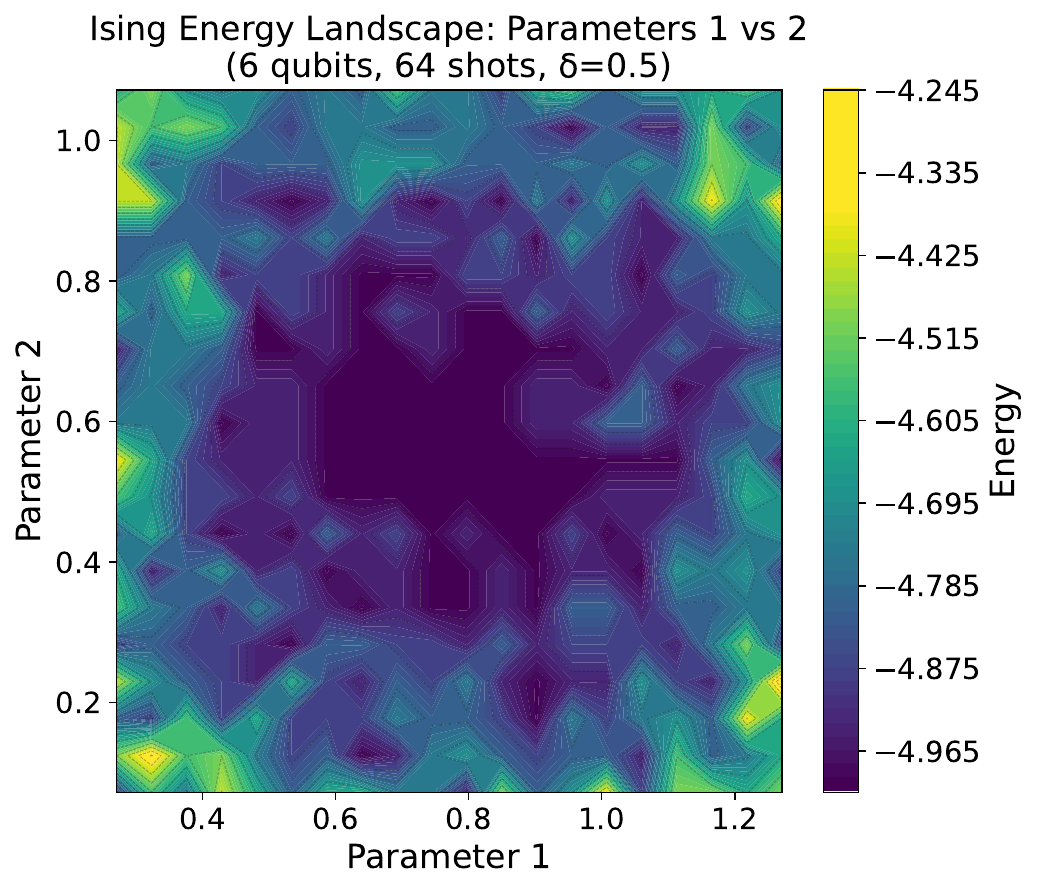}
\end{subfigure}\\[0.6em]

\begin{subfigure}[t]{0.45\textwidth}
\includegraphics[width=\linewidth]{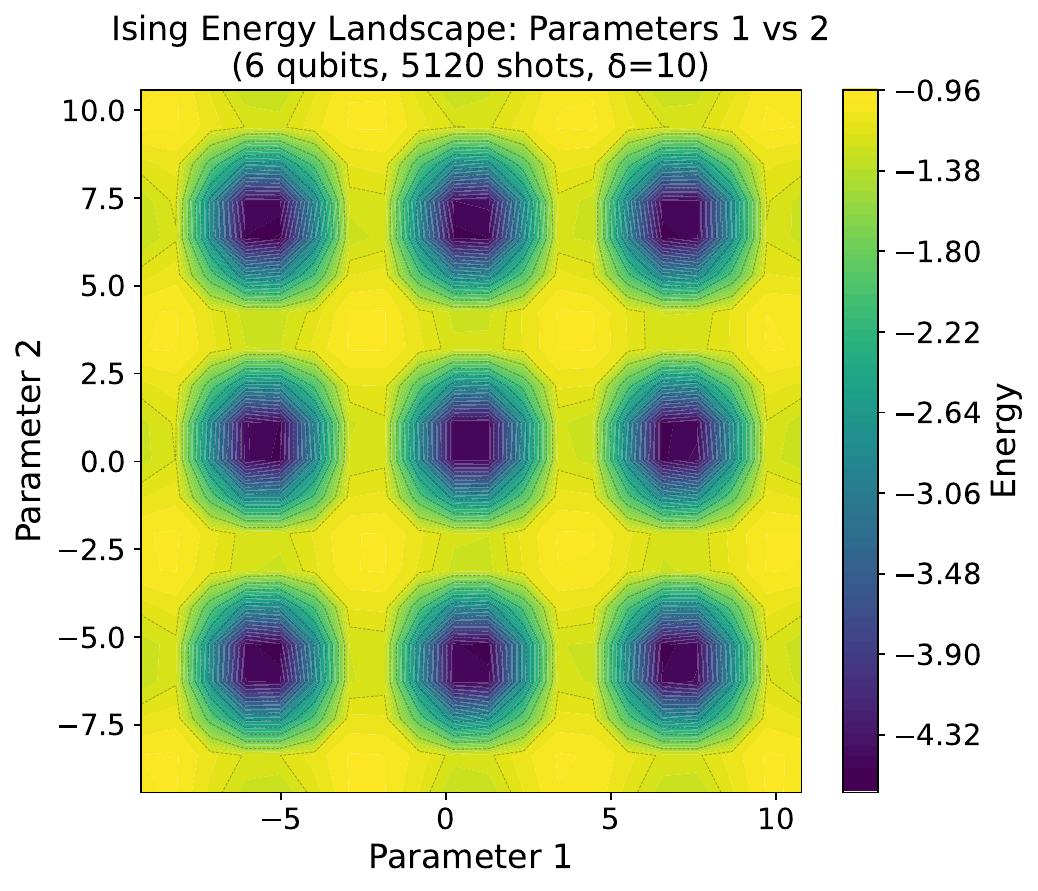}
\end{subfigure}
\begin{subfigure}[t]{0.45\textwidth}
\includegraphics[width=\linewidth]{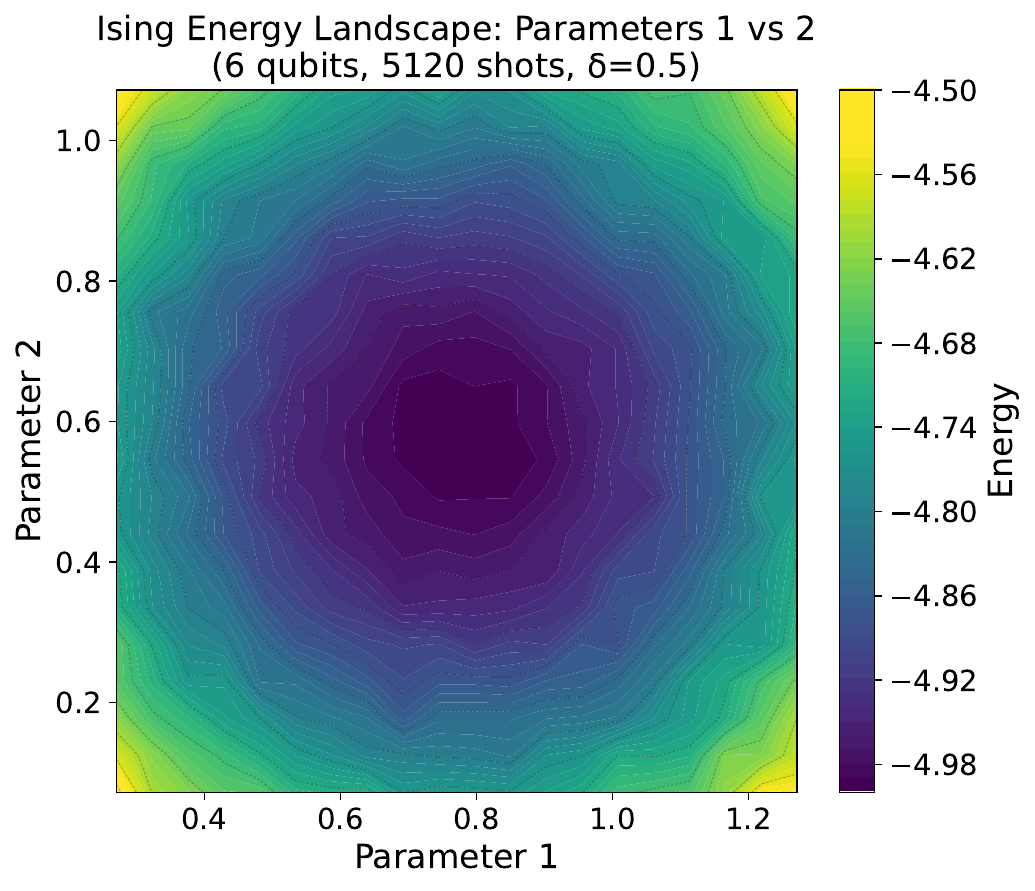}
\end{subfigure}
\caption{Energy landscapes, Ising 6 qubits, showing landscape for parameters $\theta_1$ vs $\theta_2$.}
\label{fig:landscape-6q-p1vs2}
\end{figure*}

In the Ising case the landscapes display clear periodicity induced by the ansatz and spin couplings \ref{fig:landscape-6q-p0vs1}, \ref{fig:landscape-6q-p1vs2}. In the noiseless statevector setting the local region around the minimum is smooth and approximately convex, conditions under which gradient-based local optimizers can converge reliably. Under finite-shot sampling, however, the regular contours are distorted, gradients vanish in regions, and spurious local minima appear. This is consistent with the general observation that variational landscapes can be highly sensitive to noise and prone to traps \cite{Anschuetz2022_loss_landscapes, PerezSalinas2024_information_landscapes}. As the number of shots decreases these distortions intensify, explaining the reduced efficiency of COBYLA and SPSA. Population-based algorithms such as CMA-ES and iL-SHADE, as well as swarm methods like PSO or iSOMA, are less reliant on smooth gradients and maintain robustness by sampling broadly across the space.

\begin{figure}[htpb]
\centering
\begin{tabular}{cc}
\includegraphics[width=0.48\textwidth]{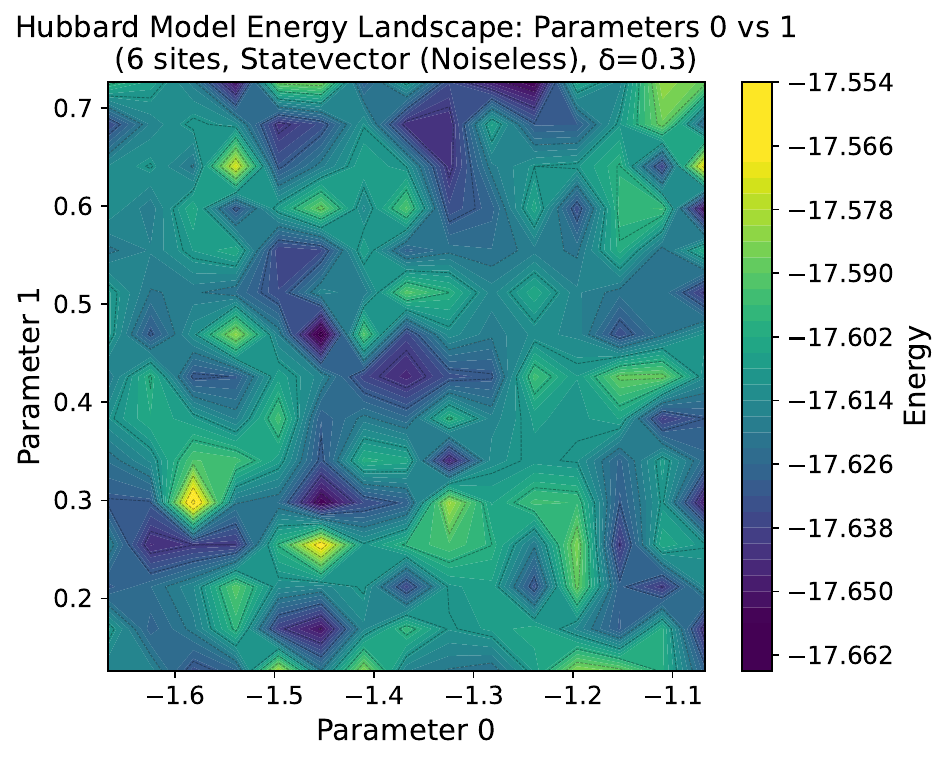} &
\includegraphics[width=0.48\textwidth]{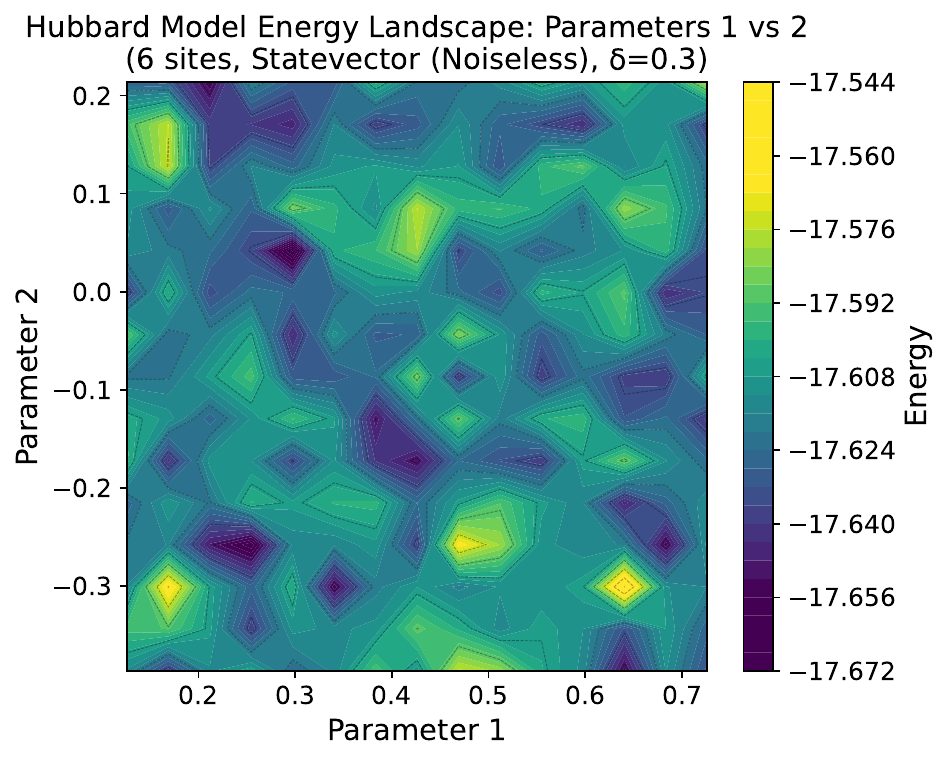} \\
\includegraphics[width=0.48\textwidth]{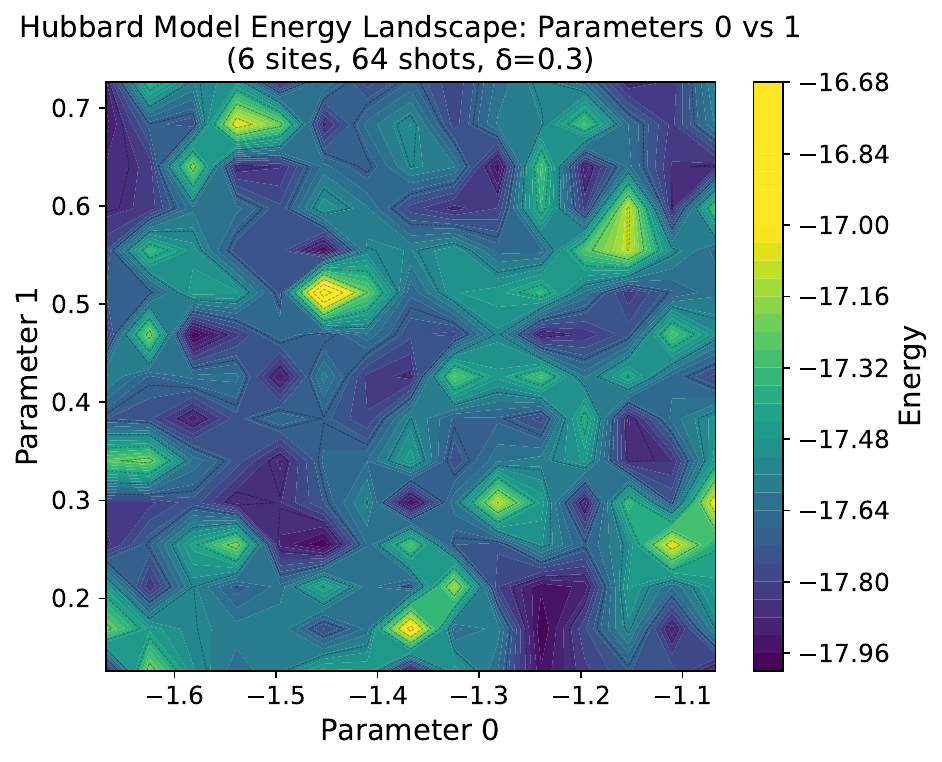} &
\includegraphics[width=0.48\textwidth]{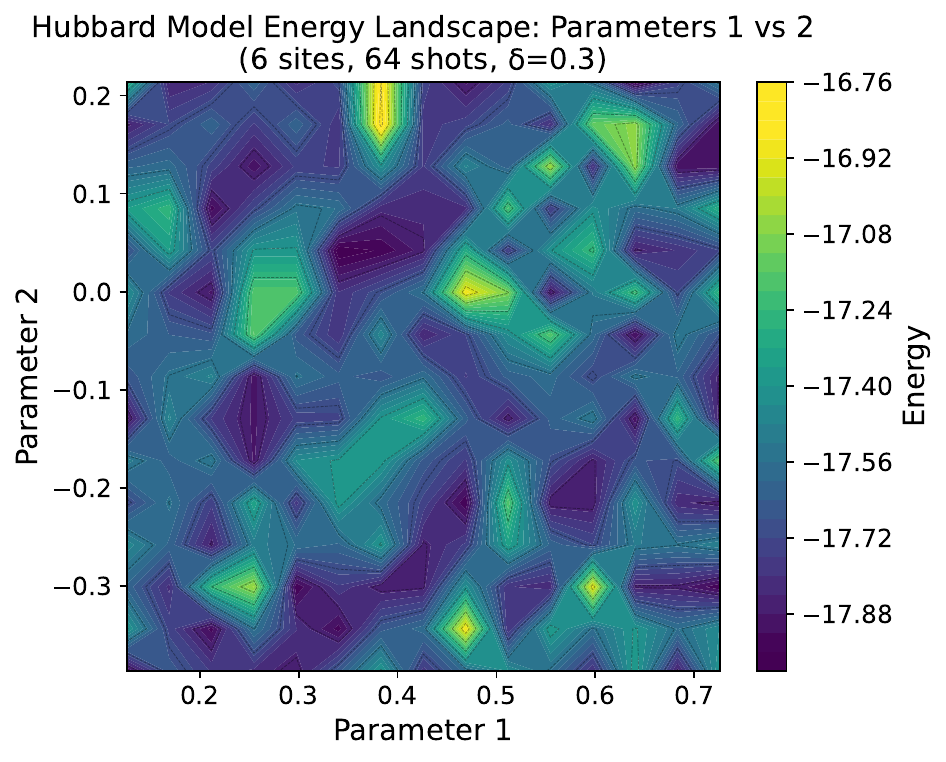} \\
\includegraphics[width=0.48\textwidth]{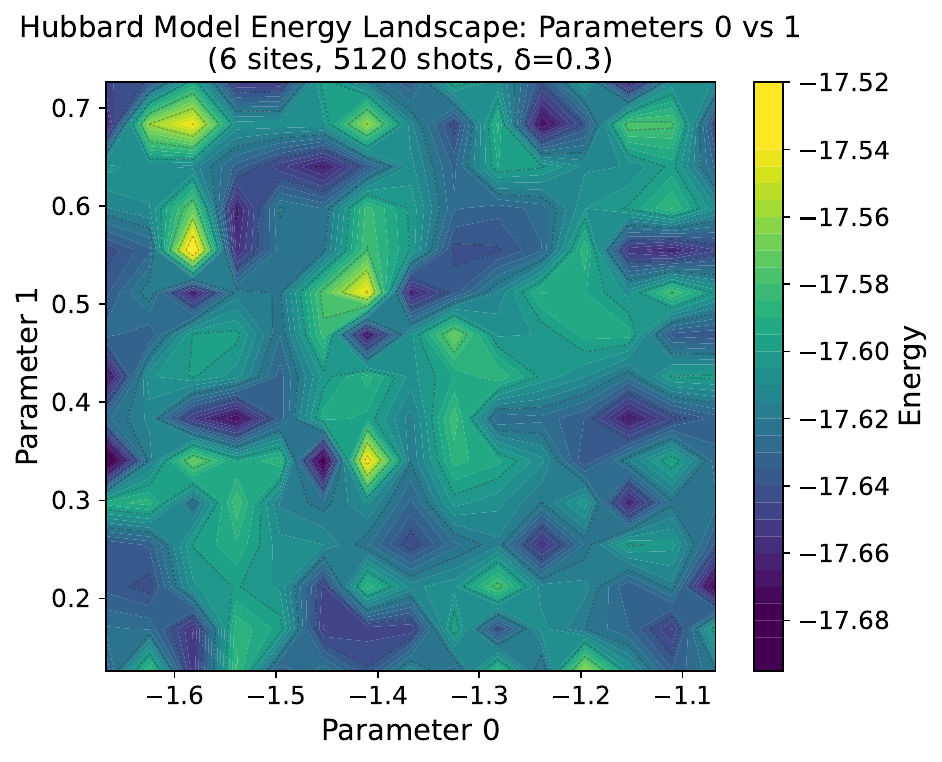} &
\includegraphics[width=0.48\textwidth]{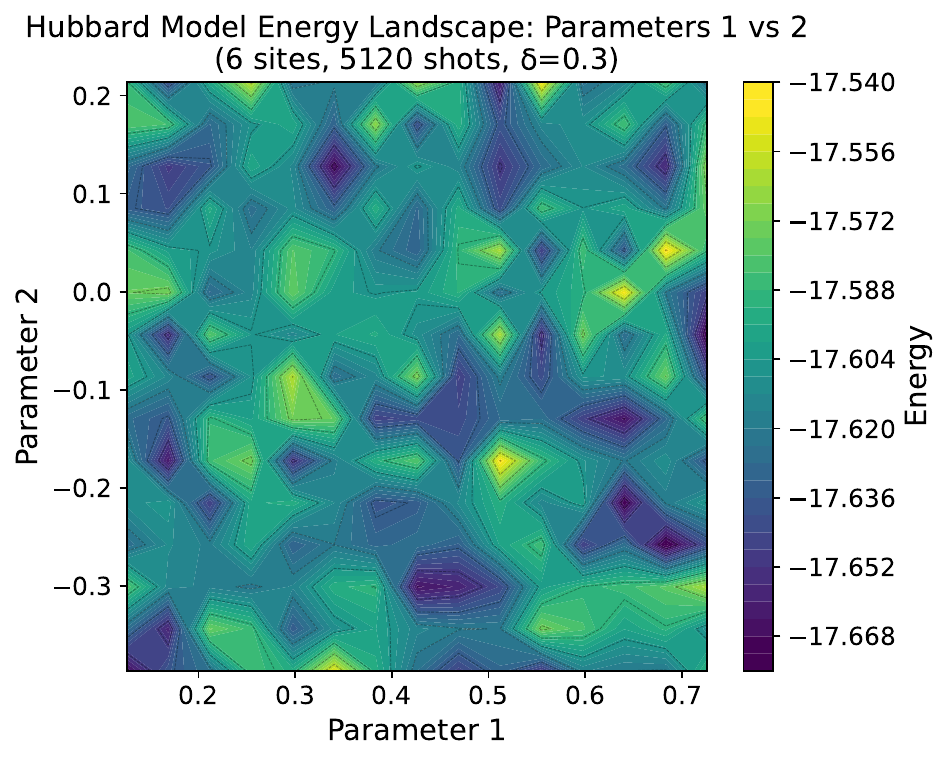} \\
\end{tabular}
\caption{Energy landscapes, Hubbard 6 site, 192 parameters, showing landscape for parameters $\theta_1$ vs $\theta_2$.}
\label{fig:landscape-hubbard}
\end{figure}

The Hubbard model presents a different picture \ref{fig:landscape-hubbard}. Its energy landscape is highly correlated and nonconvex, with many local minima scattered irregularly across the parameter space. This reflects the intrinsic difficulty of strongly correlated electron systems and aligns with benchmarks of the Hubbard model as a hard test for numerical algorithms \cite{leblanc2015solutions}. In our slicing approach, only two parameters are varied while the remaining 190 are fixed at their optimal values, so the energies in the slice remain close to the global optimum (around $-17$). The flatness of these slices masks the ruggedness of the full space, but still indicates the presence of irregular local structure. Local methods are easily trapped, while global metaheuristics are better suited to escape shallow minima and continue exploration.

These visualizations confirm that optimizer performance is dictated by landscape geometry. Periodicity and convexity can favor local descent in noiseless conditions, but noise quickly erodes this advantage. Strongly correlated models such as the Hubbard case provide inherently rugged surfaces where population-based methods are required. Similar conclusions have been reached in studies of variational energy landscapes that link circuit structure and curvature to optimizer success \cite{Kim2022_quantum_energy_landscape}. The scans with different $\delta$ values illustrate the complementary perspectives of global periodicity and local basin shape, providing context for the benchmarking results presented in the following sections.

\FloatBarrier

\FloatBarrier

\FloatBarrier

\FloatBarrier

\section{Results}
\label{sec:results}
\subsection{Local Optimizers and Limitations in Noisy Environments}

Local optimizers such as SPSA (Simultaneous Perturbation Stochastic Approximation) and COBYLA (Constrained Optimization BY Linear Approximation) are widely used in VQE due to their low computational cost and ease of implementation. In noiseless simulations, their behavior matches the landscape structure observed in Sec.~\ref{sec:landscape}: around the minimum, the Ising model displays smooth and convex basins where gradient-based descent can converge effectively. However, once sampling noise is introduced, the regular contours are distorted, gradients vanish over wide regions, and spurious minima appear. Under these conditions, COBYLA and SPSA lose reliability, frequently converging to shallow local minima or plateau regions instead of the global optimum.

\begin{figure}
\centering
\includegraphics[width=1\linewidth]{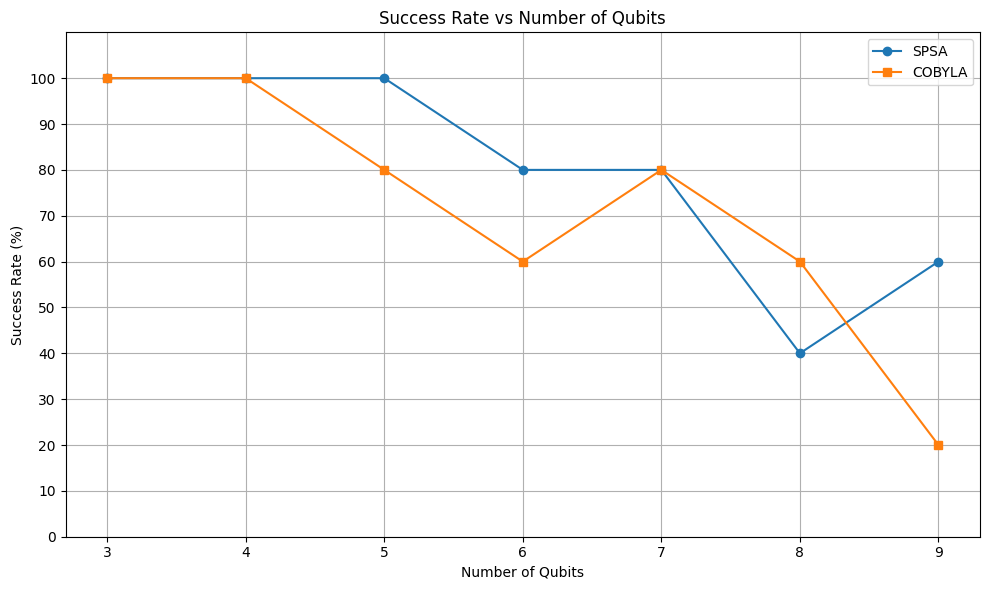}
\caption{Performance of local optimizers (COBYLA and SPSA) showing success rate decline as qubit and parameter counts increase. The SR drops sharply for COBYLA, achieving only a 20\% success rate for reaching a global minimum within a broad tolerance, while SPSA performs slightly better at approximately 50\%. The success rate measures if optimizer reached global minimum within $\delta \leq 10^{-1}$ tol.}
\label{fig}
\end{figure}

Our benchmarks (Fig.~\ref{fig}) quantify this degradation. The success rate (SR) of COBYLA falls to roughly 20\% for larger qubit counts, while SPSA maintains only about 50\%. Both rates are far below the accuracy demands of quantum chemistry applications, where tolerances as strict as $10^{-4}$ are required. The visualization of the Hubbard model further illustrates why: its rugged, nonconvex surface provides few stable gradients, making local methods prone to premature convergence regardless of noise level.

These observations indicate a structural limitation: local optimizers depend on smooth gradients that vanish or mislead in noisy, high-dimensional VQE landscapes. In contrast, population-based metaheuristics are less reliant on gradient information and can exploit broader sampling to escape distorted or flat regions, motivating their systematic evaluation in the following sections.

\subsection{Results and analysis of metaheuristic algorithms for VQE optimization}

The visualization study in Sec.~\ref{sec:landscape} already indicated that noise reshapes the optimization surface in ways that undermine local descent. Periodic and convex basins that would allow COBYLA or SPSA to succeed in noiseless conditions collapse into distorted, flat, or rugged regions once finite shots are introduced. To test which algorithms can maintain performance in such landscapes, we followed the three-phase evaluation procedure outlined in Sec.~\ref{sec:experiment_design}. This section presents the results, beginning with an initial screening across more than fifty metaheuristics, followed by scaling tests on the Ising model, and concluding with convergence on the Hubbard benchmark.

\subsection{Phase 1: Initial Screening}

The first phase assessed whether each algorithm could reach the global minimum of the 5-qubit Ising Hamiltonian within a tolerance of $10^{-1}$ in the presence of sampling noise. Algorithms were advanced to the second phase if they achieved convergence in at least one of the five runs.

In the bio-inspired group, only Symbiotic Organisms Search, Tree Physiology Optimization, Seagull Optimization, and Whale Optimization Algorithm passed, with SOS and TPO converging the fastest. Most other bio-based methods stalled under noise.  

Evolutionary algorithms performed best overall. Covariance Matrix Adaptation Evolution Strategy (CMA-ES), Differential Evolution (DE), Flower Pollination Algorithm, and advanced DE variants such as SHADE, HyDE, and iL-SHADE all succeeded. CMA-ES and iL-SHADE emerged as the most consistent, reflecting their adaptability to noisy and multimodal landscapes.  

Among human-inspired optimizers, Improved Brain Storm Optimization, Imperialist Competitive Algorithm, Life Choice Optimization, and Forensic-Based Investigation Optimization progressed, with ICA showing the highest efficiency. Most others failed to handle noise reliably.  

The math-based category was weaker: only the Runge–Kutta optimizer and Harmony Search advanced, with RUN showing efficient convergence. Harmony Search, although originating in music-inspired heuristics, belongs here in practice due to its mathematical formulation, and it achieved robust results.  

From physics-based algorithms, Fick’s Law Algorithm, Nuclear Reaction Optimization, and simulated annealing (Cauchy, Boltzmann, and fast schedules) passed. These methods showed either rapid convergence or stable robustness, consistent with their stochastic exploration mechanisms.  

Swarm-based methods were more fragile. Only Particle Swarm Optimization and the improved Self-Organizing Migrating Algorithm passed, with iSOMA outperforming PSO.  

Finally, in the system-based category, only Water Cycle Algorithm advanced, showing steady convergence under noise.  

Across categories, evolutionary algorithms were the most successful group. Their global sampling and parameter adaptation gave them clear resilience to noise-induced distortions, consistent with the landscape analysis where broad exploration avoids being misled by vanishing gradients. Physics-based and human-based methods provided a few additional candidates, while most bio-based, swarm-based, and system-based methods struggled to overcome noise effects.

\subsection{Phase 2: Function evaluation comparison on the Ising model}

In the second phase, we compared the performance of selected metaheuristics on the Ising Hamiltonian (no external field) with system sizes ranging from 3 to 9 qubits. The focus was on the number of function evaluations (FEs) needed to reach the global minimum within a tolerance of $10^{-1}$ under finite-shot noise. A higher tolerance was necessary because many algorithms could not converge to stricter levels in the presence of noise. Runs that stagnated in local minima or failed to improve over several hundred thousand evaluations are marked by ``\textemdash '' in Table~\ref{tab:algorithm_performance_sorted}.

A general trend is immediately visible: as qubit count increases, most optimizers require sharply more evaluations or fail entirely. This scaling reflects the growth in parameter space and the distortion of the energy landscape under noise. The visualization analysis in Sec.~\ref{sec:landscape} predicted exactly this: smooth convex basins at small scale allow local improvement, but as dimensionality grows, periodicity and noise create flat regions and spurious minima that trap or mislead algorithms relying on narrow search.

CMA-ES was the clear leader across all qubit sizes, requiring the fewest evaluations and showing relatively mild growth with system size. Its adaptive covariance mechanism appears to exploit the periodic structure without being destabilized by local distortions. iL-SHADE also performed strongly, though with a higher baseline cost. Unlike CMA-ES, its performance did not deteriorate severely at 9 qubits, confirming its robustness for larger systems.  

Simulated Annealing with a Cauchy distribution converged efficiently for small and mid-sized systems but required an order of magnitude more evaluations by 9 qubits. This reflects the advantage of its broad exploration early in optimization, but also its slower refinement in high-dimensional landscapes. Harmony Search exhibited good efficiency at small sizes, but its evaluation count grew steeply after 6 qubits, suggesting sensitivity to ruggedness.  

Differential Evolution variants showed mixed results. The classical DE/best/1/bin performed acceptably up to 6–7 qubits but escalated to impractically high FEs at 9 qubits. DE/best/1/exp and DE/rand1 collapsed even earlier, confirming the importance of adaptation strategies like those in iL-SHADE.  

Swarm-based approaches such as iSOMA and PSO performed well initially but degraded rapidly with dimension. iSOMA reached nearly 34,000 evaluations at 9 qubits, while PSO stagnated beyond 7 qubits. The landscapes explain this decline: swarm methods rely on smooth fitness gradients across the population, which vanish under noise and in flat regions.  

Several algorithms failed entirely for larger systems. Imperialist Competitive Algorithm and Life Choice Optimization converged at small sizes but diverged at 7 qubits or more. Virus Colony Search, Ant Lion Optimizer, and HyDE variants were among the weakest performers, requiring extreme evaluations or stagnating early. These failures illustrate the vulnerability of methods that depend heavily on exploitation dynamics when confronted with noisy, high-dimensional periodic surfaces.  

Taken together, the table highlights a separation between a small group of noise-robust, scalable optimizers—CMA-ES, iL-SHADE, and to a lesser extent SA Cauchy—and a large set of algorithms that collapse once dimensionality or noise dominates the landscape. This outcome supports the visualization insight: only strategies capable of broad exploration and dynamic adaptation can overcome the distortions and flatness that appear as system size increases.

The next phase tests whether these conclusions hold on the more complex Hubbard model, where rugged correlations rather than periodicity dominate, and the parameter count rises to 192.

\FloatBarrier
\begin{table}[ht]
\centering
\caption{Algorithm performance comparison across different qubit sizes (sorted by overall performance). Each algorithm was tested in five separate runs, and the function evaluations (FEs) were averaged for each qubit count. If an algorithm became stuck in local minima and did not improve over several hundred thousand FEs, it was denoted by "\textemdash ".}
\small
\renewcommand{\arraystretch}{0.85}
\begin{tabular}{|l|r|r|r|r|r|r|r|}
\hline
\textbf{Algorithm} & \textbf{3Q} & \textbf{4Q} & \textbf{5Q} & \textbf{6Q} & \textbf{7Q} & \textbf{8Q} & \textbf{9Q} \\
\hline
CMA-ES & 750 & 1200 & 1500 & 1800 & 2280 & 2700 & 3200 \\
SA Cauchy & 751 & 1801 & 3500 & 4201 & 7000 & 8001 & 9451 \\
iL-SHADE & 1035 & 2039 & 3274 & 3333 & 4368 & 4374 & 6559 \\
HS & 356 & 1241 & 1740 & 1726 & 5046 & 7066 & 10586 \\
DE best1bin & 948 & 1520 & 4460 & 8112 & 11676 & 14720 & 17136 \\
iSOMA & 1357 & 3245 & 5552 & 15263 & 22115 & 28144 & 33899 \\
SOS & 1866 & 2720 & 4880 & 13440 & 15320 & 32000 & 37843 \\
DE best1exp & 3028 & 8706 & 21213 & 47232 & 56132 & 64032 & 72036 \\
LCO & 3000 & 8000 & 14000 & 28650 & 43500 & 50000 & \textemdash  \\
ICA & 1100 & 2350 & 4850 & 7650 & 62000 & \textemdash  & \textemdash  \\
GA & 5150 & 19050 & 21050 & 22050 & 38850 & 81150 & 75750 \\
DE/rand1 & 4100 & 17300 & 35000 & 30000 & 35000 & 150000 & \textemdash  \\
CRO & 788 & 5875 & 22000 & 30000 & 24000 & \textemdash  & \textemdash  \\
FPA & 1500 & 7000 & 12400 & 30000 & 50000 & \textemdash  & \textemdash  \\
SA fast & 7201 & 10401 & 23001 & 33601 & 38501 & 45601 & 51301 \\
IBSO & 17000 & 38000 & 38000 & 38000 & 41000 & 42000 & 44000 \\
BBO & 1800 & 9500 & 16800 & 20000 & 31300 & 39680 & \textemdash  \\
WOA & 1800 & 4033 & 17666 & 57500 & \textemdash  & \textemdash  & \textemdash  \\
SOMA T3A & 3775 & 10102 & 16847 & 57882 & \textemdash  & \textemdash  & \textemdash  \\
HyDE & 8450 & 38000 & 41200 & 60000 & \textemdash  & \textemdash  & \textemdash  \\
FBIO & 4000 & 28400 & 48000 & 65000 & 154000 & \textemdash  & \textemdash  \\
HyDE-DF & 1600 & 23000 & 23000 & 80000 & 72000 & 100000 & \textemdash  \\
PSO & 6200 & 6480 & 18400 & 80000 & 130000 & \textemdash  & \textemdash  \\
VCS & 32000 & 50000 & 76000 & 80000 & 120000 & \textemdash  & \textemdash  \\
SA Boltzmann & 4801 & 35601 & 50501 & 86401 & \textemdash  & \textemdash  & \textemdash  \\
ALO & 51500 & 100000 & 100000 & 200000 & 300000 & 300000 & \textemdash  \\
TPO & 17000 & 26000 & 26000 & \textemdash  & \textemdash  & \textemdash  & \textemdash  \\
SOA & 10000 & 13000 & 26000 & \textemdash  & \textemdash  & \textemdash  & \textemdash  \\
\hline
\end{tabular}
\label{tab:algorithm_performance_sorted}
\end{table}
\FloatBarrier

\subsection{Phase 3: Convergence on the Hubbard Model}
\raggedbottom

\begin{figure}
    \centering
    \includegraphics[width=1\linewidth]{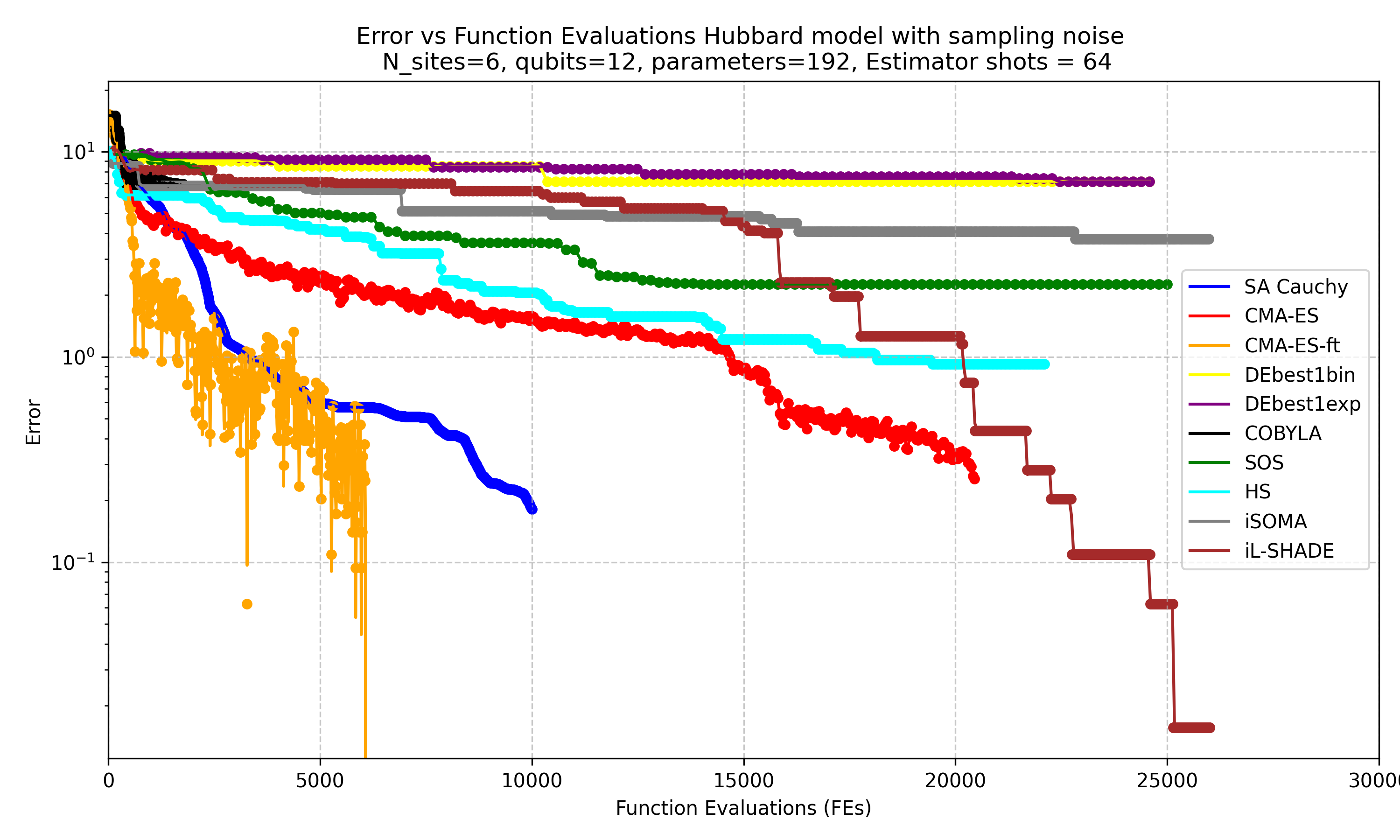}
    \caption{Convergence of various optimization algorithms on a 6-site Hubbard model with 192 parameters and 64 shots per measurement. The reduced number of shots introduces increased noise, creating a more rugged energy landscape and making convergence more challenging. Fine tuned CMA-ES (orange) demonstrated the fastest and most consistent convergence to the global minimum, outperforming SA with multivariate Cauchy (blue) and other CMA-ES runs (red, crimson, salmon). iL-SHADE (brown) showed competitive performance, although slightly slower, but reaching lower error. Harmony Search (cyan) provided reasonable results, while SciPy implementations of Differential Evolution (yellow, purple) and iSOMA (gray) failed to effectively navigate the noisy optimization landscape. The substantial noise levels emphasized the robustness of algorithms like CMA-ES and SA Cauchy in finding near-optimal solutions in challenging conditions.}
    \label{hubbard_64}
\end{figure}

Figure \ref{hubbard_64} illustrates the convergence behavior of various optimization methods for estimating expectation values with only 64 shots per measurement. Under these conditions, the results are highly susceptible to sampling noise, which introduces fluctuations in the estimated energy values. This noise creates challenges in accurately navigating the energy landscape, as low-shot estimations can obscure the true gradient direction, making it difficult for optimization algorithms to converge to the global minimum. For all optimizers, we show the mean of 5 independent runs.

Compared to the Ising benchmark, the Hubbard model presents a markedly harsher landscape. Correlated fermionic structure, sign constraints, and stronger entanglement produce rugged, nonconvex regions with many local traps. As our visualization already suggested, these traps can both hinder convergence and, at finite depth, bias where apparent “critical” behavior emerges. Recent studies show that locally trapped variational states still encode phase information and can support finite-depth extrapolation to locate transitions, but they remain hard for classical optimizers to escape \cite{cao_traps_vqa}. By contrast, work on Differential Evolution in VQE reports strong performance on Ising and 1D Hubbard under exact simulations without shot noise; our noisy setting explains the divergence in outcomes and motivates algorithms explicitly robust to stochastic flattening and spurious minima \cite{faildo2023using}.

The measurements indicate that the Covariance Matrix Adaptation Evolution Strategy with fine tuning (more in Sec. \ref{appendix_params}) is the best-performing optimizer, reaching the exact global minimum with the lowest number of function evaluations. The DE variant iL-SHADE reached the exact global minimum after a higher number of FEs. In contrast, other methods required tens of thousands of FEs to achieve even relatively high tolerances. Simulated Annealing using a multivariate Cauchy distribution had fast convergence, but struggled to find global minimum in lower tolerances. Harmony Search and Symbiotic Organisms Search followed in effectiveness.

Interestingly, well-regarded algorithms like Differential Evolution and the improved Self-Organizing Migrating Algorithm, which typically excel at optimizing classical functions, performed poorly on VQE landscapes. The DE algorithm, implemented from SciPy \cite{2020SciPy-NMeth}, was unable to complete the predetermined number of function evaluations, as it terminated early based on a stagnation convergence criterion: failing to achieve improvements over a large number of FEs. This premature termination highlights the challenges that even robust optimizers face when applied to VQE problems, where the high-dimensional and complex energy landscapes often lead to stagnation. Such inconsistencies between optimizers are intriguing, as some methods significantly outperform others depending on the specific variationally parameterized circuits used. The underlying reasons why certain optimizers adapt more effectively to VQE landscapes remain an open question, suggesting an area ripe for further investigation into the interactions between algorithm behavior and quantum circuit structures.

\begin{figure}[H]
    \centering
    \includegraphics[width=1\linewidth]{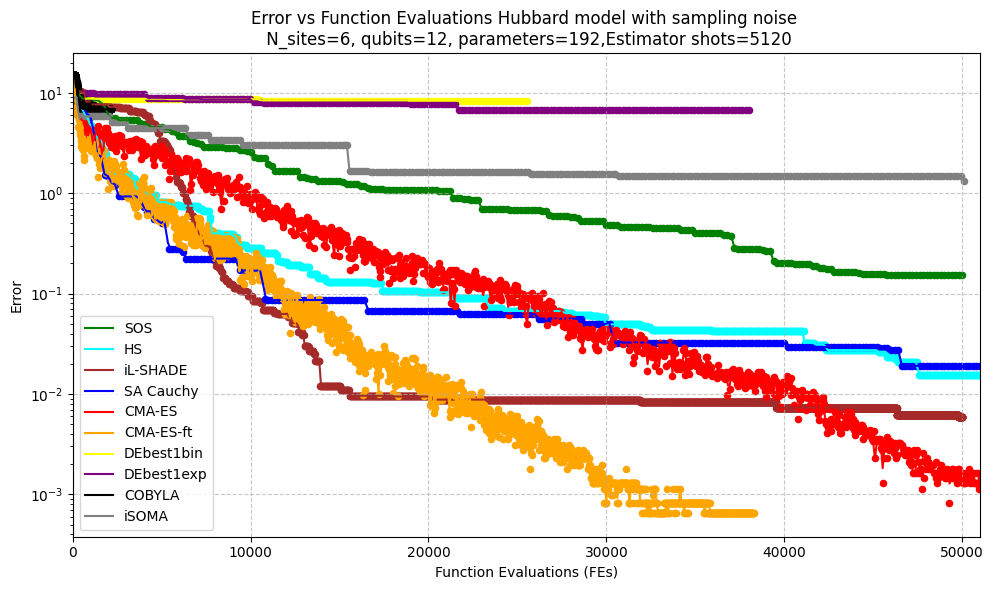}
    \caption{Convergence of various optimization algorithms on a 6-site Hubbard model with 192 parameters and 5120 shots per measurement. As the number of shots increases, noise is reduced, leading to a smoother energy landscape. Fine tuned CMA-ES (orange) and iL-SHADE (brown) exhibited the most promising convergence rates, with CMA-ES achieving the global minimum with the lowest error. SA Cauchy and Harmony Search (cyan) displayed similar convergence patterns, albeit slightly less effective. The SciPy implementations of Differential Evolution (DEbest1bin and DEbest1exp) failed to converge to a meaningful solution, highlighting their limitations in noisy landscapes. iSOMA (gray) converged poorly, stagnating at higher error values. The extended 6-site Hubbard model, used as the optimization target in the VQE, introduced additional complexity due to its high-dimensional parameter space and the inclusion of sampling noise. This significantly increased the challenge of identifying the ground state energy, as the noise created a highly rugged and deceptive energy landscape.}

    \label{hubbard_61024}
\end{figure}

Figure \ref{hubbard_61024} presents the convergence behavior when the number of shots increases to 5120. In this scenario, the trend remains the same, with fine tuned CMA-ES still leading, but Harmony Search begins to show behavior similar to SA with the Cauchy distribution, narrowing the performance gap. This suggests that while CMA-ES and SA maintain their advantage with higher accuracy, the increased reliability of energy estimations reduces the relative benefit of SA’s broader exploration capabilities.

However, our research also highlighted the significant limitations of local gradient-based optimizers, such as SPSA and COBYLA, especially in noisy conditions. These optimizers frequently terminated prematurely, often becoming trapped in high-valued local minima, which emphasizes the nonconvex nature of the Hubbard model's energy landscape.

\subsection{Discussion of results and comparison to previous studies}

A central finding of our study is the divergence between results obtained under noiseless simulations, which dominate much of the existing literature, and results that include realistic sampling noise. Many prior works establish the strength of metaheuristics such as Differential Evolution (DE) in exact simulations of Ising or Hubbard models. Our experiments reveal that this picture changes substantially once stochastic noise is introduced, reshaping the landscape and altering algorithm performance.

\paragraph{Using Differential Evolution to avoid local minima in Variational Quantum Algorithms.}
A recent study \cite{faildo2023using} reports that DE with binomial and exponential crossover outperforms local optimizers in noiseless VQE simulations, exploiting its exploratory ability to escape local minima. Hybrid DE schemes were proposed as especially promising. Our findings qualify this result: in noisy conditions, standard DE as implemented in SciPy degrades severely, often terminating prematurely due to stagnation. By contrast, advanced DE variants such as iL-SHADE remain competitive even on the 192-parameter Hubbard problem, highlighting the necessity of adaptive strategies for robustness. These observations echo performance gaps also seen in optimization challenges, where SHADE-type algorithms consistently outperform classical DE baselines.

\paragraph{Performance of DE variants in CEC competitions.}
DE-based methods are well established as state-of-the-art in real-parameter constrained optimization, as shown in the CEC 2017 and 2019 competitions \cite{brest2019100, lezama2019hybrid, yeh2019modified, fu2019univariate}. Our study partly confirms this reputation: iL-SHADE is among the best algorithms tested. However, the broader class of DE variants underperformed relative to CMA-ES and Simulated Annealing in noisy VQE. The discrepancy illustrates that benchmarks on deterministic functions do not fully capture the stochastic ruggedness of quantum variational landscapes.

\paragraph{Swarm intelligence and reinforcement learning.}
Work on reinforcement-learning-enhanced swarm optimizers such as iSOMA \cite{klein2024optimizing} shows strong performance in deterministic settings, competitive with leading DE variants. Our results agree that iSOMA is among the more robust swarm approaches, but under noise it lags behind CMA-ES and iL-SHADE. This confirms that while reinforcement learning can improve parameter control, stochastic flattening in VQE requires deeper algorithmic robustness.

\paragraph{Noise-aware classical optimizers.}
Prior investigations into classical optimizers for NISQ devices \cite{lavrijsen2020classical} found that gradient-based methods such as COBYLA and BFGS, excellent in noiseless settings, break down once even modest noise is introduced. Our results reproduce this effect: COBYLA and SPSA, the only gradient-based algorithms tested, consistently failed under shot noise. This matches landscape analyses that show gradients vanish or fluctuate erratically in noisy VQE.

\paragraph{Comparisons across VQE benchmarks.}
Recent work on optimization in correlated systems \cite{cao_traps_vqa} demonstrates that variational landscapes often contain shallow traps that prevent reaching the ground state, though these trapped solutions can still encode meaningful physical information. Our results for the Hubbard model align with this picture: local optimizers frequently terminate in suboptimal minima, while global methods such as CMA-ES and iL-SHADE escape more effectively. Similarly, \cite{bonet2023performance} finds that CMA-ES with careful hyperparameter tuning can outperform other optimizers under noise. Our broader benchmarking extends these conclusions: Simulated Annealing with Cauchy distribution, Harmony Search, and Symbiotic Organisms Search also achieve strong convergence, suggesting a wider set of viable noise-robust methods than previously recognized.

Taken together, these comparisons indicate that algorithm rankings from noiseless studies must be re-evaluated in noisy, large-parameter landscapes. CMA-ES emerges as consistently best-performing across conditions, with iL-SHADE and a small set of stochastic heuristics providing complementary robustness. This reinforces the importance of testing optimizers in realistic noisy environments rather than relying solely on idealized simulations.

\section*{Conclusion}

This study provided a systematic evaluation of more than fifty metaheuristic algorithms for Variational Quantum Algorithms, with an emphasis on their robustness under sampling noise. A multiphase procedure was used: an initial screening on a 5-qubit Ising model, scaling tests on Ising models up to 9 qubits, and convergence studies on the 192-parameter Hubbard model. In parallel, we introduced visualization of optimization landscapes, which clarified how algorithm behavior is shaped by geometry and noise.

The landscape analysis revealed two key features: smooth, convex basins in noiseless Ising systems that favor local descent, and rugged, noise-distorted or flat regions in both noisy Ising and Hubbard models that undermine gradient-based methods. These observations explain the systematic failure of COBYLA and SPSA once noise is introduced, and they motivate the use of global, sampling-based metaheuristics that do not depend on stable gradients.

Across benchmarks, CMA-ES emerged as the most reliable optimizer, consistently reaching the global minimum with low evaluation counts and scaling gracefully with system size. Among Differential Evolution methods, iL-SHADE was the only variant that maintained competitive performance under noise, confirming the importance of adaptive strategies. Simulated Annealing with a Cauchy distribution, Harmony Search, and Symbiotic Organisms Search also performed strongly, particularly on the Hubbard model, showing that lesser-used heuristics can provide practical robustness in VQE contexts. By contrast, many widely known metaheuristics—including PSO, GA, and WOA—collapsed beyond small system sizes, either stagnating or requiring prohibitively many evaluations.

The Hubbard model provided the most stringent test: its correlated fermionic structure generates a rugged, nonconvex surface where local methods consistently fail. Here, CMA-ES and iL-SHADE again led, with SA Cauchy and HS narrowing the gap as shot numbers increased. These results align with recent reports that VQE landscapes can trap optimizers in shallow minima while still encoding physical information, underscoring the need for noise-resilient global methods.

Taken together, our findings suggest that metaheuristic performance on noiseless benchmarks is not predictive of outcomes on noisy quantum simulations. Algorithms that excel in deterministic competitions or exact Ising/Hubbard studies often degrade sharply once sampling noise is introduced. CMA-ES, iL-SHADE, and a small set of stochastic heuristics represent robust exceptions. For future work, hybrid strategies that combine local refinement with noise-robust global exploration appear especially promising, as they could bridge the gap between efficiency and reliability in realistic VQE settings.

\section*{CRediT authorship contribution statement}
\textbf{Vojtěch Novák}: Software, Investigation, Writing – original draft. \textbf{Ivan Zelinka}: Supervision, Validation, Reviewing. \textbf{Václav Snášel:} Conceptualization, Supervision.

\section*{Acknowledgements}
The following grants are acknowledged for the financial support provided for this research: grant of SGS No. SP2024/008, VSB-Technical University of Ostrava, Czech Republic.

\section*{Declaration of Competing Interest}
The authors declare that they have no known competing financial interests or personal relationships that could have appeared to influence the work reported in this paper.

\section*{Data availability}
All simulations were conducted using a locally developed code, which is openly available for further examination and reproduction of results. The code, along with detailed Jupyter notebooks, can be accessed at \url{https://github.com/VojtechNovak/VQA_metaheuristics}. Further details are available through the corresponding author.

\section{Appendix A : details of the algorithms}
\label{sec:details}
\subsection*{Evolutionary algorithms}

\textbf{Covariance Matrix Adaptation Evolution Strategy (CMA-ES)}: An advanced evolutionary algorithm for continuous optimization that adapts the covariance matrix of the search distribution to learn the objective function's landscape. Key hyperparameters include population size (typically scaled with problem dimension) and step-size control (sigma). This adaptation enables effective exploration and exploitation of complex multimodal functions \cite{cmaes2001}.

\textbf{Coral Reef Optimization (CRO)}: Inspired by coral reef ecosystems, simulating competition for space, larval settlement, reproduction, and predation. Hyperparameters control occupied reef rate, broadcast spawning rates, asexual reproduction rates, depredation probabilities, mutation factors, and settlement trials. These parameters balance exploration through larval settlement and exploitation through local reproduction \cite{cro2014}.

\textbf{Evolutionary Programming (EP)}: A classical evolutionary algorithm emphasizing mutation and selection over recombination. Primary hyperparameters include population size and bout size for tournament selection. Particularly effective for rugged or multimodal landscapes where recombination may be less beneficial \cite{yao1999evolutionary}.

\textbf{Evolution Strategies (ES)}: A family of self-adaptive evolutionary algorithms using mutation, recombination, and selection. Key parameters include population size and offspring-to-parent ratio (lambda). The self-adaptation of strategy parameters like mutation step sizes makes ES robust for high-dimensional optimization \cite{hansen2015evolution}.

\textbf{Flower Pollination Algorithm (FPA)}: Models flower pollination through biotic and abiotic mechanisms. Hyperparameters include population size, switch probability between global and local pollination, and Levy flight multipliers. Global pollination enables exploration while local pollination provides exploitation \cite{abdel2019flower}.

\textbf{Memetic Algorithm (MA)}: Hybrid approach combining evolutionary search with local refinement. Parameters include population size, crossover and mutation probabilities, local search probability, maximum local generations, and encoding precision. This combination of global evolutionary operations and local search enhances convergence quality \cite{neri2012memetic}.

\textbf{Genetic Algorithm (GA)}: Inspired by natural selection using selection, crossover, and mutation operations. Core hyperparameters are population size, crossover probability, and mutation probability. These parameters control the balance between preserving good solutions and introducing genetic diversity \cite{mirjalili2019genetic}.

\textbf{Success-History Adaptation Differential Evolution (SHADE)}: Enhanced DE variant with self-adaptive parameter control. Hyperparameters include population size, initial mutation factor, and crossover probability means. SHADE dynamically adjusts these parameters based on success history from previous generations \cite{shade2013}.

\textbf{Improved L-SHADE (iL-SHADE)}: Advanced SHADE variant with linear population reduction. Key parameters include initial population size (typically scaled with dimension), memory size for parameter adaptation, and individual solution dimension. The population reduction strategy improves convergence efficiency.

\textbf{Hybrid Differential Evolution (HyDE)}: Combines DE with additional optimization techniques for enhanced performance. Standard DE parameters include mutation weight factor and crossover rate, with hybrid components adding local search or gradient-based refinement strategies \cite{hyde2015}.

\textbf{Differential Evolution (DE)}: Multi-particle strategy using mutation and crossover operations. Primary hyperparameters are mutation weight factor and crossover rate, with variants using different crossover schemes (binomial vs. exponential). Population size affects success rate and computational demands \cite{price2013differential}.

\subsection*{Swarm-based algorithms}

\textbf{Self-Organizing Migrating Algorithm (SOMA)}: A stochastic optimization method inspired by social dynamics of competitive and cooperative individuals. Operating through iterative Migration loops, SOMA evaluates populations and designates the best individual as Leader. Key hyperparameters include number of jumps, step size, population size, perturbation rate, and migration parameters controlling the movement toward leaders. The perturbation parameter introduces controlled randomness during migration, while path length determines traversal extent. Our implementation uses iSOMA, an advanced variant that has demonstrated superior performance over SOMA T3A \cite{skanderova2023self,zelinka2016soma,zelinka2000soma}.

\textbf{Particle Swarm Optimization (PSO)}: Inspired by collective swarm behavior in nature, PSO employs cooperative search patterns based on individual learning and swarm communication. Core hyperparameters include population size, inertia weight, and cognitive/social acceleration coefficients. The algorithm adheres to five artificial life principles: proximity (space-time computation), quality (environmental sensing), diverse response (unrestricted resource seeking), stability (consistent behavior), and adaptability (behavioral adjustment). Particles update positions and velocities based on personal best and global best information, balancing exploration and exploitation \cite{eberhart1995particle,van2001analysis}.

\textbf{Whale Optimization Algorithm (WOA)}: Models humpback whale bubble-net hunting strategy through encircling prey, bubble-net attacking, and prey searching mechanisms. Hyperparameters include hunting party size and spiral parameter controlling the logarithmic spiral movement. The algorithm transitions between exploration (random search) and exploitation (encircling behavior) phases, with the spiral updating mechanism mimicking the whales' bubble-net feeding behavior \cite{WOA2016}.

\textbf{Artificial Bee Colony (ABC)}: Simulates honey bee foraging behavior using three bee types: employed bees (exploit known food sources), onlooker bees (select sources based on quality), and scout bees (explore new areas). Key parameters include number of food sources, employed bees, onlooker bees, and abandonment limit. The algorithm balances exploration through scout bees and exploitation through employed/onlooker bee cooperation, with information sharing determining source selection \cite{ABC2009}.

\textbf{Ant Lion Optimizer (ALO)}: Inspired by ant lion predatory behavior, simulating trap construction, prey entrapment, and capture processes. Primary hyperparameter is colony size, with the algorithm modeling random walks of ants, trap building by ant lions, and prey capture mechanisms. The hunting simulation provides effective exploration through random walks and exploitation through trap-based convergence \cite{ALO2015}.

\textbf{Elephant Herding Optimization (EHO)}: Based on elephant social structure and matriarchal leadership, modeling clan-based organization and herding behavior. Hyperparameters control population organization and clan dynamics. The algorithm employs clan updating operations reflecting elephant social coordination and separating operations mimicking natural herd splitting, balancing exploration and exploitation through social hierarchy simulation \cite{EHO2015}.

\textbf{Harris Hawks Optimization (HHO)}: Models cooperative hunting strategies of Harris' hawks through multiple hunting phases. Key parameters control exploration intensity, transition dynamics, and exploitation strategies. The algorithm simulates surprise pounce tactics and collaborative hunting behavior, transitioning from exploration (perching and random search) to exploitation (surprise pounce with different attack strategies) based on prey energy levels \cite{HHO2019}.

\subsection*{Bio-Based Algorithms}

This section provides an overview of various bio-based algorithms used for optimization. Each algorithm is inspired by natural phenomena and biological processes.

\textbf{Brown Bear Optimization Algorithm (BBOA)}: Inspired by brown bear foraging behavior and survival strategies. Primary hyperparameter is population size. The algorithm models territorial behavior, seasonal migration patterns, and food-seeking strategies to balance exploration of new territories with exploitation of known resource-rich areas \cite{bbobear2023}.

\textbf{Slime Mould Algorithm (SMA)}: Mimics adaptive foraging behavior of slime moulds navigating toward food sources. Key parameters include probability threshold and population size. The algorithm simulates oscillation patterns, shrinking mechanisms, and position updates based on food quality, enabling effective exploration through random search and exploitation through convergence toward optimal food sources \cite{sma2020}.

\textbf{Biogeography-Based Optimization (BBO)}: Based on species migration patterns between habitats, modeling biogeographic distribution principles. Hyperparameters include mutation probability, number of elite solutions, and population size. The algorithm uses habitat suitability indices, migration rates, and mutation operators to share information between solutions, simulating species migration and habitat colonization \cite{bbo2008}.

\textbf{Barnacles Mating Optimizer (BMO)}: Models reproductive behavior and settlement strategies of barnacles. Key parameters include population size and threshold values for reproductive success. The algorithm simulates larval dispersal, settlement site selection, and mating proximity constraints to balance exploration through dispersal and exploitation through local reproductive success \cite{bmo2018}.

\textbf{Earthworm Optimization Algorithm (EOA)}: Simulates earthworm movement and soil-foraging strategies including reproduction, crossover, mutation, and cooling mechanisms. Hyperparameters include population size, crossover and mutation probabilities, and temperature-based cooling factors. The algorithm models soil navigation, reproduction cycles, and environmental adaptation for effective landscape exploration \cite{eoa2015}.

\textbf{Satin Bowerbird Optimizer (SBO)}: Inspired by male bowerbird mating behavior and bower construction for mate attraction. Parameters include population size, mutation probability, and proportion of space width. The algorithm simulates bower building, decoration arrangement, and mate selection processes, balancing exploration through territory establishment and exploitation through mate attraction strategies \cite{sbo2017}.

\textbf{Seagull Optimization Algorithm (SOA)}: Models seagull migration patterns and hunting behavior including collision avoidance and spiral movement. Key parameters are population size and frequency control. The algorithm simulates flocking behavior, spiral diving for prey capture, and social coordination during migration, providing exploration through migration and exploitation through coordinated hunting \cite{soa2018}.

\textbf{Symbiotic Organisms Search (SOS)}: Inspired by ecological symbiotic relationships including mutualism, commensalism, and parasitism. Primary hyperparameter is population size. The algorithm models beneficial, neutral, and competitive interactions between organisms to create cooperative search strategies that enhance both exploration and exploitation through symbiotic partnerships \cite{sos2014}.

\textbf{Tree Physiology Optimization (TPO)}: Simulates physiological processes of trees including photosynthesis, transpiration, and growth mechanisms. Parameters include population size and growth control factors (alpha, beta, theta). The algorithm models nutrient absorption, energy conversion, and adaptive growth responses to environmental conditions \cite{tpo2018}.

\textbf{Tunicate Swarm Algorithm (TSA)}: Based on collective movement and foraging strategies of marine tunicates. Primary hyperparameter is population size. The algorithm simulates swarm coordination, jet propulsion movement, and filter-feeding behavior to navigate marine environments, combining exploration through swarm movement with exploitation through coordinated foraging \cite{tsa2020}.

\textbf{Virus Colony Search (VCS)}: Models viral propagation and infection mechanisms including replication and host selection. Key parameters include population size, selection ratio (lambda), and infection strength (sigma). The algorithm simulates viral spread patterns, host adaptation, and replication strategies to explore solution spaces through infection propagation and exploit through successful replication \cite{vcs2015}.

\textbf{Wildebeest Herd Optimization (WHO)}: Mimics migratory behavior and social coordination of wildebeest herds during seasonal migration. Complex parameter set includes exploration/exploitation steps, learning rates, movement probabilities, and distance controls. The algorithm models herd formation, leadership dynamics, and coordinated movement across landscapes, balancing exploration through migration routes with exploitation through herd cooperation \cite{who2019}.

\subsection*{Physics-Based Algorithms}

This subsection provides an overview of various physics-based algorithms used for optimization. Each algorithm is inspired by physical phenomena and principles.

\textbf{Simulated Annealing (SA):} \cite{kirkpatrick1983optimization} stands as a concise yet powerful technique renowned for its efficacy in solving both single and multiple objective optimization problems with notable efficiency gains. Drawing inspiration from the process of metal cooling and annealing in thermodynamics, SA operates by mimicking the gradual crystallization of liquid metals to their lowest energy state. The algorithm uses key hyperparameters including maximum and minimum temperatures that control the cooling schedule, the length of temperature range determining the number of cooling steps, and a maximum stay counter limiting iterations at each temperature level. By navigating through the landscape of potential solutions with a probabilistic approach akin to the annealing process, SA effectively bypasses local optima to converge towards the global optimum.

\textbf{Atom Search Optimization (ASO):} \cite{aso2018} The ASO algorithm is inspired by atomic motion under Newtonian mechanics and the Lennard-Jones potential. It mimics atomic behaviors under attractive and repulsive forces to explore the search space. Key hyperparameters include alpha (depth weight) controlling exploitation intensity, beta (multiplier weight) affecting force calculations, and population size determining the number of atoms in the system.

\textbf{Archimedes Optimization Algorithm (ArchOA):} \cite{archoa2020} Based on Archimedes' principle describing buoyant forces on submerged objects, ArchOA simulates floating and sinking behavior in fluids. The algorithm uses scaling factors (c1-c4) to control various force components, along with maximum and minimum acceleration limits that constrain particle movement dynamics, enabling effective balance between exploration and exploitation.

\textbf{Chernobyl Disaster Optimizer (CDO):} \cite{cdo2023} Inspired by radioactive particle dispersion following the Chernobyl disaster, CDO models contamination spread using diffusion principles. The algorithm primarily relies on population size as its main hyperparameter, with the dispersion dynamics inherently managing exploration through the simulation of wide-ranging particle impact.

\textbf{Electromagnetic Field Optimization (EFO):} \cite{abedinpourshotorban2016electromagnetic} Models charged particle behavior in electromagnetic fields using attraction and repulsion forces. Key hyperparameters include reproduction rate (analogous to mutation), population selection rate (crossover-like parameter), and field proportions controlling positive and negative field influences on particle interactions.

\textbf{Equilibrium Optimizer (EO):} \cite{eo2019} Inspired by mass balance principles from control volume theory, EO simulates dynamic system equilibrium. The algorithm uses population size as its primary hyperparameter, with equilibrium dynamics naturally balancing exploration and exploitation through the inherent system stability mechanisms.

\textbf{Energy Valley Optimizer (EVO):} \cite{evo2022} Based on energy valley concepts in potential energy surfaces, EVO mimics particle movement toward lowest energy states. The algorithm uses population size to control the number of particles exploring energy valleys, leveraging natural energy minimization tendencies.

\textbf{Fick's Law Algorithm (FLA):} \cite{fla} Inspired by diffusion laws describing particle flux from high to low concentration areas. The algorithm employs multiple control factors (C1-C5) governing different aspects of the diffusion process and a distance factor managing spatial relationships between particles during concentration-driven movement.

\textbf{Henry Gas Solubility Optimization (HGSO):} \cite{hgso2019} Based on Henry's law describing gas solubility in liquids under varying pressure. The algorithm uses population size and number of clusters to organize gas particles, with clustering controlling the formation of gas groups that simulate dissolution and release dynamics.

\textbf{Multi-Verse Optimizer (MVO):} \cite{mvo} Inspired by multiple universe concepts in physics, MVO simulates inter-universe information exchange. Key hyperparameters include wormhole existence probability ranges (minimum and maximum) that control the likelihood of solution transfer between universes, enabling global optimum location through multi-dimensional exploration.

\textbf{Nuclear Reaction Optimization (NRO):} \cite{nro2019} Inspired by nuclear reactions, including fusion and fission processes, NRO simulates these reactions to explore and exploit the search space. By modeling energy release and particle interactions during nuclear reactions, NRO effectively searches for optimal solutions, reflecting the dynamic processes of nuclear physics.

\textbf{RIME Optimization (RIME):} \cite{rime2023} Based on rime ice formation phenomena, RIME simulates ice accumulation and sublimation processes. The algorithm uses population size and a soft-rime parameter that controls the ice formation characteristics, balancing exploration and exploitation through phase change dynamics.

\textbf{Tug of War Optimization (TWO):} \cite{two} Inspired by the competitive tug of war game, TWO models force balance between opposing teams. The algorithm primarily uses population size to determine the number of participants, with competitive dynamics naturally emerging from the strategic pulling mechanisms.

\textbf{Wind Driven Optimization (WDO):} \cite{wdo} Inspired by atmospheric air mass movement, WDO simulates wind flow dynamics. The algorithm employs multiple hyperparameters including RT coefficient, gravitational constant controlling gravitational influence, alpha (update constant), Coriolis effect parameter, and maximum speed limit, collectively modeling pressure gradients and atmospheric forces.

\subsection*{Human-Based Algorithms}

This subsection provides an overview of various human-based algorithms used for optimization. Each algorithm is inspired by human behavior and social phenomena.

\textbf{Battle Royale Optimization (BRO):} \cite{bro} Inspired by survival-of-the-fittest concepts in battle royale games, BRO uses competitive elimination until one remains. The algorithm employs population size and a threshold parameter (dead threshold) that determines elimination criteria, simulating natural selection through competitive exclusion.

\textbf{Improved Brain Storm Optimization (BSO):} \cite{ibso} Enhanced brainstorming-inspired algorithm that prevents premature convergence. Key hyperparameters include population size, number of clusters organizing idea groups, and probability parameters (p1-p4) controlling local/global search splits, idea development phases, and cluster center weighting preferences.

\textbf{Culture Algorithm (CA):} \cite{ca} Inspired by cultural evolution in human societies, CA combines genetic algorithms with cultural learning. The algorithm uses population size and an accepted rate parameter controlling the probability of accepting cultural knowledge, guiding solution evolution through belief space knowledge transmission.

\textbf{Coronavirus Herd Immunity Optimization (CHIO):} \cite{chio} Models virus spread and containment within populations, simulating herd immunity phenomena. Key hyperparameters include population size, basic reproduction rate controlling infection spread, and maximum age limiting infected case duration, balancing exploration through epidemic dynamics.

\textbf{Forensic-Based Investigation Optimization (FBIO):} \cite{fbio} Mimics forensic investigation processes in criminal cases, using evidence collection and hypothesis validation. The algorithm primarily relies on population size, with investigation dynamics inherently managing exploration through systematic evidence-based search strategies.

\textbf{Gaining Sharing Knowledge-based Algorithm (GSKA):} \cite{gska} Inspired by knowledge sharing in human societies, GSKA simulates social learning interactions. Key hyperparameters include best percent (proportion of top candidates) and knowledge ratio influencing learning intensity, enabling efficient exploration through collaborative knowledge exchange.

\textbf{Heap-based Optimizer (HBO):} \cite{hbo2020} Inspired by heap data structures, HBO uses heap properties for search space prioritization. The algorithm employs population size and degree parameter controlling the corporate rank hierarchy levels, effectively managing exploration and exploitation balance.

\textbf{Human Conception Optimizer (HCO):} \cite{hco2022} Models biological human conception processes, simulating sperm cell competition. Key hyperparameters include population size, weight factors for probability and velocity updates, and acceleration coefficients controlling particle movement dynamics during competitive fertilization simulation.

\textbf{Imperialist Competitive Algorithm (ICA):} \cite{ica2007} Inspired by imperialism and empire competition, ICA models colonial assimilation processes. The algorithm uses empire count, assimilation coefficient, revolution probability and rate, revolution step size, and colonies coefficient (zeta) to manage imperialist-colony dynamics and competitive empire interactions.

\textbf{Life Choice-based Optimization (LCO):} \cite{lco2019} Inspired by human life decision-making processes, LCO simulates choice evaluation and selection. The algorithm employs population size and a coefficient factor (r1) that influences decision-making dynamics during life choice exploration.

\textbf{Queuing Search Algorithm (QSA):} \cite{qsa} Models queue dynamics to manage search space exploration, using queuing principles to balance exploration and exploitation. The algorithm primarily relies on population size with inherent queue management controlling search dynamics.

\textbf{Search And Rescue Optimization (SARO):} \cite{saro} Simulates coordinated search and rescue operations, modeling systematic search strategies. Key hyperparameters include population size, social effect controlling social influence strength, and maximum unsuccessful searches limiting failed search attempts.

\textbf{Student Psychology Based Optimization (SPBO):} \cite{spbo2020} Inspired by student psychological behaviors in learning environments, SPBO models educational problem-solving strategies. The algorithm primarily uses population size with student psychology dynamics naturally managing search space navigation.

\textbf{Social Ski-Driver Optimization (SSDO):} \cite{ssdo2019} Inspired by ski driver social behaviors, SSDO simulates social interactions and decision-making. The algorithm relies on population size with social dynamics inherently controlling exploration and exploitation balance.

\textbf{Teaching Learning-based Optimization (TLO):} \cite{tlo2012} Inspired by educational teaching-learning processes, TLO models teacher-learner interactions for iterative solution improvement. The algorithm primarily uses population size with educational dynamics naturally guiding solution quality enhancement.

\textbf{Teamwork Optimization Algorithm (TOA):} \cite{toa2021} Inspired by teamwork and collaboration principles, TOA simulates team member coordination and cooperation. The algorithm relies on population size with collaborative teamwork dynamics naturally managing search space exploration.

\textbf{War Strategy Optimization (WARSO):} \cite{wso} Inspired by warfare strategies, WARSO models tactical and strategic military decision-making. Key hyperparameters include population size and switching probability (rr) controlling strategy update probability, enabling strategic navigation through war-like competitive dynamics.

\subsection*{Math-Based Algorithms}

This subsection provides an overview of various math-based algorithms used for optimization. Each algorithm is inspired by mathematical concepts and techniques.

\textbf{Arithmetic Optimization Algorithm (AOA):} \cite{aoa} Inspired by basic arithmetic operations, AOA uses addition, subtraction, multiplication, and division for iterative solution refinement. Key hyperparameters include alpha (sensitive exploitation parameter), control parameter (miu) adjusting search processes, and Math Optimizer Accelerated range parameters (moa\_min/moa\_max) controlling acceleration bounds.

\textbf{Cross-Entropy Method (CEM):} \cite{cem2005} A probabilistic optimization technique estimating optimal solutions through distribution sampling and parameter updates. The algorithm uses population size, number of best solutions for next generation selection, and weight factor (alpha) controlling normal distribution mean and standard deviation calculations.

\textbf{Chaos Game Optimization (CGO):} \cite{cgo2020} Inspired by fractal-generating chaos games, CGO uses chaotic maps to explore search spaces. The algorithm primarily relies on population size, with chaotic system ergodicity naturally providing randomness for local minima avoidance.

\textbf{Circle Search Algorithm (CSA):} \cite{circlesa2022} Inspired by geometric circle properties, CSA explores through iterative circle expansion and contraction. Key hyperparameters include population size and convergence factor controlling the convergence rate, balancing exploration and exploitation through geometric dynamics.

\textbf{Gradient-Based Optimizer (GBO):} \cite{gbo} Utilizes gradient information for search space navigation following steepest descent principles. The algorithm employs population size, probability parameter controlling search probability, and beta range parameters (beta\_min/beta\_max) as fixed algorithmic constants.

\textbf{Hill Climbing (HC):} \cite{hc} A local search algorithm moving toward increasing/decreasing values until reaching peaks/valleys. Key hyperparameters include population size and neighbor size determining the number of neighboring solutions to consider during local search operations.

\textbf{Weighted Mean of Vectors (INFO):} \cite{info2022} Inspired by weighted average mathematical concepts, INFO combines multiple candidate solutions through weighted mean computation. The algorithm primarily uses population size with weighted averaging dynamics naturally guiding iterative solution refinement.

\textbf{Pareto-like Sequential Sampling (PSS):} \cite{pss} Inspired by Pareto principles, PSS uses sequential sampling strategies focusing on balanced exploration-exploitation regions. Key hyperparameters include population size, acceptance rate for solution acceptance probability, and sampling method choice (Latin-Hypercube or Monte Carlo) for solution generation.

\textbf{Runge Kutta Optimizer (RUN):} \cite{run} Inspired by Runge-Kutta differential equation methods, RUN applies numerical techniques to optimization through objective function derivative approximation. The algorithm primarily relies on population size with numerical method dynamics naturally managing iterative improvement.

\textbf{Sine Cosine Algorithm (SCA):} \cite{sca} Inspired by trigonometric sine and cosine functions, SCA leverages periodic properties for diverse search patterns. The algorithm primarily uses population size with trigonometric function periodicity naturally ensuring comprehensive search space coverage.

\textbf{Success History Intelligent Optimizer (SHIO):} \cite{shio} Uses historical success information to guide search processes, dynamically adjusting strategies based on past performance. The algorithm primarily relies on population size with success history dynamics naturally enhancing optimal solution finding capabilities.

\textbf{Tabu Search (TS):} \cite{tabu} A metaheuristic using memory structures to avoid cycles and prevent revisiting explored solutions. Key hyperparameters include population size, tabu size controlling maximum tabu list size, neighbor size for neighboring solution consideration, and perturbation scale controlling solution modification intensity.

\subsection*{Music-Based Algorithms}

This subsection provides an overview of optimization algorithms inspired by music.

\textbf{Harmony Search (HS):} \cite{yang2009harmony} Inspired by musician improvisation processes, HS models harmony creation through pitch adjustment for optimal combinations. Key hyperparameters include Harmony Memory Consideration Rate controlling the probability of selecting values from harmony memory versus generating new ones, and Pitch Adjustment Rate controlling the probability of adjusting selected values within harmony memory, balancing exploration and exploitation.

\subsection*{System-Based Algorithms}

This subsection provides an overview of various system-based algorithms used for optimization. Each algorithm is inspired by natural or artificial systems.

\textbf{Artificial Ecosystem-based Optimization (AEO):} \cite{zhao2020artificial} Inspired by energy flow and organism interactions in ecosystems, AEO models production, consumption, and decomposition processes. The algorithm primarily uses population size to control the number of organisms, with ecosystem dynamics naturally mimicking balance and sustainability found in natural systems.

\textbf{Germinal Center Optimization (GCO):} \cite{villasenor2018germinal} Inspired by immune system germinal center biological processes, GCO simulates B-cell affinity maturation. Key hyperparameters include crossover rate controlling crossover occurrence probability and weighting factor influencing crossover operations, enabling iterative solution improvement through biological selection, mutation, and recombination analogies.

\textbf{Water Cycle Algorithm (WCA):} \cite{wca2012} Inspired by natural water cycles including precipitation, evaporation, and stream flow, WCA models water droplet movement and interaction. Key hyperparameters include number of rivers plus sea controlling water body distribution, weighting coefficient for river and sea weighting, and evaporation condition constant controlling evaporation thresholds, balancing exploration through natural hydrological processes.

\section{APPENDIX B: Algorithm Parameters}
\label{appendix_params}
This appendix lists the hyperparameters used for each optimization algorithm in our experiments. Parameters are presented in the format: \texttt{parameter\_name=value [typical\_range]}, where \texttt{value} represents our chosen setting and \texttt{[typical\_range]} indicates commonly used values in the literature. Algorithms are grouped by their underlying optimization paradigm. In the preliminary testing phase (Phase 1), we used the default parameter settings for all optimizers as listed here. By default we refer to the values provided in the respective software packages or commonly adopted in the literature. These defaults are designed to balance exploration and exploitation and are generally considered robust starting points.

We additionally fine tuned the \texttt{CMA-ES-ft} optimizer. Our choice to focus on CMA-ES was pragmatic: preliminary experiments showed that it consistently achieved the best and fastest convergence among all tested algorithms, making it the most promising candidate for hyperparameter optimization. Fine tuning was carried out with \texttt{IRACE}, which iteratively updates its candidate generation model based on surviving configurations and generates new configurations to race against the elites until the evaluation budget is depleted. The hyperparameters tuned were the population size $\in [15,120]$ and the initial step-size $\sigma_0 \in [0.1,1.0]$. We allocated 150 evaluations of the hyperparameters, and the surviving configurations returned by \texttt{IRACE} were selected as the optimal settings for \texttt{CMA-ES-ft}. In contrast, the other promising candidate, \texttt{iL-SHADE}, did not require additional tuning due to its adaptive nature: the algorithm employs success-history based parameter adaptation together with a linear population size reduction scheme, which makes it largely self-adjusting without the need for external hyperparameter optimization. We note that systematic hyperparameter tuning could also improve the performance of other algorithms. However, we expect the general performance hierarchy observed in our experiments, particularly the poor scaling of swarm-based methods in high dimensions, to remain unchanged.

\footnotesize
\setlength{\parskip}{2pt}

\textbf{Evolutionary Algorithms:}
CRO: pop\_size=50, po=0.4 [0.2-0.5], Fb=0.9 [0.6-0.9], Fa=0.1 [0.05-0.3], Fd=0.1 [0.05-0.5], Pd=0.5 [0.1-0.7], GCR=0.1 [0.05-0.2], gamma\_min=0.02 [0.01-0.1], gamma\_max=0.2 [0.1-0.5], n\_trials=5 [2-10]. %
EP: pop\_size=50, bout\_size=0.05 [0.05-0.2]. %
ES: pop\_size=50, lambda=0.75 [0.5-1.0]. %
FPA: pop\_size=50, p\_s=0.8 [0.5-0.95], levy\_multiplier=0.2 [0.0001-1000]. %
MA: pop\_size=50, pc=0.85 [0.7-0.95], pm=0.15 [0.05-0.3], p\_local=0.5 [0.3-0.7], max\_local\_gens=10 [5-25], bits\_per\_param=4. %
GA: pop\_size=50, pc=0.9 [0.7-0.95], pm=0.05 [0.05-0.3]. %
SHADE: pop\_size=100, miu\_f=0.5 [0.4-0.6], miu\_cr=0.5 [0.4-0.6]. %
iL-SHADE: pop\_size=12*dim, memory\_size=6. %
CMA-ES: pop\_size=5*N, sigma=0.5. %
CMA-ES-ft: pop\_size=24, sigma=0.798. %
HyDE/DE: F\_weight=0.5, F\_CR=0.6. %
MPEDE: lambdas=[0.2,0.2,0.2,0.4], ng=20, c=0.1 [0-1], p=0.04 (0-1].

\textbf{Swarm Algorithms:}
iSOMA/SOMA/SOMA T3A: N\_jump=10, Step=0.11, PopSize=40, prt=0.1, m=30, n=20, s=3. %
PSO: pop=40, w=0.8, c1=0.5, c2=0.5.

\textbf{Bio-inspired Algorithms:}
SMA: z=0.03 [0.01-0.1], pop\_size=100. %
BBOA: pop\_size=50. %
BBO: pm=[0.01-0.2], n\_elites=[2-5], pop\_size=100. %
BMO: threshold=4, pop\_size=200. %
EOA: pop\_size=50, pc=0.9 [0.5-0.95], pm=0.01 [0.01-0.2], n\_best=2, alpha=0.98 [0.8-0.99], beta=0.9 [0.8-1.0], gamma=0.9 [0.8-0.99]. %
SBO: pop\_size=200, alpha=0.7 [0.5-2.0], pm=0.05 [0.01-0.2], psw=0.02 [0.01-0.1]. %
SOA: pop\_size=75, fc=4 [1-5]. %
SOS: pop\_size=50. %
TPO: pop\_size=50, alpha=0.3, beta=50 [10-50], theta=0.9 [0.5-0.9]. %
TSA: pop\_size=200. %
VCS: pop\_size=50, lambda=0.5 [0.2-0.5], sigma=0.3 [0.1-2.0]. %
WHO: pop\_size=50, n\_explore\_step=3 [2-4], n\_exploit\_step=3 [2-4], eta=0.15 [0.05-0.5], p\_hi=0.9 [0.7-0.95], local\_alpha=0.9 [0.5-0.9], local\_beta=0.3 [0.1-0.5], global\_alpha=0.2 [0.1-0.5], global\_beta=0.8, delta\_w=2.0 [1.0-2.0], delta\_c=2.0 [1.0-2.0]. %
ABCO: food\_sources=30, employed\_bees=15, outlookers\_bees=10, limit=1. %
ALO: colony\_size=100. %
AOA: size=500, alpha=0.5, mu=5. %
WOA: hunting\_party=15, spiral\_parameter=2.

\textbf{Physics-based Algorithms:}
ASO: pop\_size=100, alpha=50, beta=0.2. %
ArchOA: pop\_size=50, c1=2 [1-2], c2=5 [2,4,6], c3=2 [1-2], c4=0.5 [0.5-1], acc\_max=0.9, acc\_min=0.1. %
CDO: pop\_size=100. %
EFO: pop\_size=50, r\_rate=0.3 [0.1-0.6], ps\_rate=0.85 [0.5-0.95], p\_field=0.1, n\_field=0.45. %
EO: pop\_size=50. %
EVO: po\_psize=50. %
FLA: pop\_size=50, C1=0.5, C2=2.0, C3=0.1, C4=0.2, C5=2.0, DD=0.01. %
HGSO: pop\_size=50, n\_clusters=3 [2-10]. %
MVO: pop\_size=50, wep\_min=0.2 [0.05-0.3], wep\_max=1.0 [0.75-1.0]. %
RIME: pop\_size=50, sr=5.0. %
TWO: pop\_size=50. %
WDO: pop\_size=50, RT=3, g\_c=0.2 [0.1-0.5], alpha=0.4 [0.3-0.8], c\_e=0.4 [0.1-0.9], max\_v=0.3 [0.1-0.9]. %
SAFast/SABoltzmann/SACauchy: T\_max=100, T\_min=1e-7, L=300, max\_stay\_counter=150.

\textbf{Human-based Algorithms:}
BRO: pop\_size=100, threshold=3 [2-5]. %
BSO: pop\_size=50, m\_clusters=5 [3-10], p1=0.25, p2=0.5, p3=0.75, p4=0.6 [0.4-0.6]. %
CA: pop\_size=50, accepted\_rate=0.15 [0.1-0.5]. %
CHIO: pop\_size=50, brr=0.15 [0.05-0.2], max\_age=100 [5-20]. %
FBIO: pop\_size=50. %
GSKA: pb=0.1 [0.1-0.5], kr=0.9 [0.5-0.9]. %
HBO: pop\_size=50, degree=3 [2-4]. %
HCO: pop\_size=50, wfp=0.65, wfv=0.05, c1=1.4, c2=1.4. %
ICA: pop\_size=50, empire\_count=5 [3-10], assimilation\_coeff=1.5 [1.0-3.0], revolution\_prob=0.05 [0.01-0.1], revolution\_rate=0.1 [0.05-0.2], revolution\_step\_size=0.1 [0.05-0.2], zeta=0.1 [0.05-0.2]. %
LCO: pop\_size=50, r1=2.35 [1.5-4]. %
QSA: pop\_size=50. %
SARO: pop\_size=50, se=0.5 [0.3-0.8], mu=15. %
SPBO: pop\_size=50. %
SSDO: pop\_size=50. %
TLO: pop\_size=50. %
TOA: pop\_size=50. %
WARSO: pop\_size=50, rr=0.1 [0.1-0.9].

\textbf{Math-based Algorithms:}
AOA: pop\_size=50, alpha=5 [3-8], miu=0.5 [0.3-1.0], moa\_min=0.2, moa\_max=0.9. %
CEM: pop\_size=50, n\_best=20, alpha=0.7. %
CGO: pop\_size=50. %
CSA: pop\_size=50, c\_factor=0.8. %
GBO: pop\_size=50, pr=0.5 [0.2-0.8], beta\_min=0.2, beta\_max=1.2. %
HC: pop\_size=50, neighbour\_size=50 [2-1000]. %
INFO: pop\_size=50. %
PSS: pop\_size=50, acceptance\_rate=0.8 [0.7-0.96], sampling\_method=LHS. %
RUN: pop\_size=50. %
SCA: pop\_size=50. %
SHIO: pop\_size=50. %
TS: pop\_size=200, tabu\_size=5 [5-10], neighbour\_size=20 [5-100], perturbation\_scale=0.05 [0.01-1].

\textbf{Music-based Algorithms:}
HS: c\_r=0.95 [0.1-0.5], pa\_r=0.05 [0.3-0.8].

\textbf{System-based Algorithms:}
AEO: pop\_size=50. %
GCO: cr=0.5, wf=1.5. %
WCA: nsr=4, wc=2, dmax=1e-6.

\bibliographystyle{apsrev4-2}
\bibliography{bibsample}

\end{document}